\documentclass[useAMS,usenatbib]{mn2e}
\usepackage{times}
\usepackage{amssymb}
\usepackage{graphicx}
\usepackage{rotating}
\usepackage{lscape}
\usepackage{multirow}
\usepackage{fixltx2e}

\def\totaltarget{381} 
\def\totaltargetH{313} 
\def\totaltargetJ{68} 

\def\totalHafivesig{193} 
\def\totalHafivesigH{152} 

\def\totalmissingHa{188} 
\def\missingHaduetolowsig{101} 
\def\missingleft{87} 

\def\totalHaNIItwosig{64}
\def\totalHaOIIItwosig{67}
\def\totalHaHbtwosig{13}
\def\totalBPT{3}

\begin{document}

%
%
%
%



\title[The HiZELS FMR at $z\sim1-1.5$]{A fundamental metallicity relation for galaxies at ${\bf {\emph z}=0.84-1.47}$ from HiZELS} 

\author[J.P. Stott et al.]{John P. Stott$^{1}$\thanks{E-mail: j.p.stott@durham.ac.uk}, David Sobral$^{2}$, Richard Bower$^{1}$, Ian Smail$^{1}$, Philip N. Best$^{3}$, \newauthor   Yuichi Matsuda$^{4,5}$,  Masao Hayashi$^{6}$, James E. Geach$^{7}$, Tadayuki Kodama$^{4,8}$\\
\\
$^{1}$ Institute for Computational Cosmology, Durham University, South Road, Durham, DH1 3LE, UK\\
$^{2}$ Leiden Observatory, Leiden University, P.O. Box 9513, NL-2300 RA Leiden, The Netherlands\\
$^{3}$ SUPA, Institute for Astronomy, Royal Observatory of Edinburgh, Blackford Hill, Edinburgh, EH9 3HJ, UK\\
$^{4}$ National Astronomical Observatory of Japan, 2-21-1 Osawa, Mitaka, Tokyo 181-8588, Japan\\
$^{5}$ The Graduate University for Advanced Studies (SOKENDAI), 2-21-1 Osawa, Mitaka, Tokyo 181-8588, Japan\\
$^{6}$ Institute for Cosmic Ray Research, The University of Tokyo, Kashiwa, 277-8582, Japan\\
$^{7}$ Centre for Astrophysics Research, Science \& Technology Research Institute, University of Hertfordshire, Hatfield, AL10 9AB, UK \\
$^{8}$ Subaru Telescope, National Astronomical Observatory of Japan, 650 North AÕohoku Place, Hilo, HI 96720, USA\\
}

\date{}

\pagerange{\pageref{firstpage}--\pageref{lastpage}} \pubyear{2013}

\maketitle

\label{firstpage}

\begin{abstract}

We obtained Subaru FMOS observations of $\rm H\alpha$ emitting galaxies selected from the HiZELS narrow-band survey, to investigate the relationship between stellar mass, metallicity and star-formation rate at $z=0.84-1.47$, for comparison with the Fundamental Metallicity Relation seen at low redshift. Our findings demonstrate, for the first time with a homogeneously selected sample, that a relationship exists for typical star-forming galaxies at $z\sim1-1.5$ and that it is surprisingly similar to that seen locally. Therefore, star-forming galaxies at $z\sim1-1.5$ are no less metal abundant than galaxies of similar mass and star-formation rate (SFR) at $z\sim0.1$, contrary to claims from some earlier studies. We conclude that the bulk of the metal enrichment for this star-forming galaxy population takes place in the 4\,Gyr before $z\sim1.5$. We fit a new mass-metallicity-SFR plane to our data which is consistent with other high redshift studies. However, there is some evidence that the mass-metallicity component of this high redshift plane is flattened, at all SFR, compared with $z\sim0.1$, suggesting that processes such as star-formation driven winds, thought to remove enriched gas from low mass halos, are yet to have as large an impact at this early epoch. The negative slope of the SFR-metallicity relation from this new plane is consistent with the picture that the elevation in the SFR of typical galaxies at $z\gtrsim1$ is fuelled by the inflow of metal-poor gas and not major merging. 

\end{abstract}


\section{Introduction}

Gas phase metallicity is a key parameter to constrain the current stage of galaxy evolution because it reflects the result of past star-forming activity and the history of both gas inflow and outflow of the system. The presence of a stellar mass-metallicity relation (sometimes referred to as the MZR) for star-forming galaxies was first observed by \cite{lequeux1979},  with more massive galaxies possessing higher gas metallicity, and this is now well established at $z\sim0.1$ (e.g., \citealt{tremonti2004,kewley2008}). The origin of this relation is thought to be due to massive galaxies, with their deeper gravitational potential wells, being better able to hold onto their enriched gas in the presence of strong star formation driven winds, than their less massive counterparts (an idea first proposed by \citealt{larson1974} in relation to elliptical galaxies; see also \citealt{arimoto1987}). 

Tracking the evolution of the mass-metallicity relation of star-forming galaxies with redshift should help to illuminate the physical processes responsible for the peak in the volume averaged star formation rate (SFR) for galaxies at $z=1-2$ and its subsequent downturn to the present day (e.g. \citealt{lilly1996,madau1996,sobral2013}). The mass-metallicity relation of star-forming galaxies is generally observed to shift to lower metallicity with increasing redshift, relative to the $z=0.1$ relation of \cite{tremonti2004} (e.g. \citealt{savaglio2005,erb2006, maiolino2008,lamareille2009,perezm2009,yabe2012,zahid2013}). This would suggest that galaxy metallicity evolves strongly from $z\gtrsim1$ to the present. However, instead of constituting an evolution, it has been suggested by \cite{mannucci2010} that these results are due to the high redshift samples having significantly higher SFRs. This is because the observed high redshift galaxies are often UV selected from flux limited samples with high SFRs, even compared to the elevated specific SFR of typical galaxies at these epochs. The UV selection tends to bias against metal-rich/dusty galaxies while being complete for very metal-poor galaxies. In fact, \cite{mannucci2010} find that these high redshift, high SFR galaxies sit on the same `Fundamental Metallicity Relation' (FMR, see also \citealt{laralopez2010,laralopez2013,cresci2012}), a plane in mass-metallicity-SFR, as that of $z\sim0.1$ galaxies from the Sloan Digital Sky Survey (SDSS, \citealt{sdss}). An extension to this FMR for low mass galaxies was published in \cite{mannucci2011}. The shape of the FMR is to first order a manifestation of the positive correlation of the metallicity with stellar mass at fixed SFR (i.e. the mass-metallicity relation) and a negative correlation with SFR at fixed stellar mass. The latter relation is thought to be because the inflowing fuel for star formation has its origins in the relatively metal-poor inter-galactic medium (IGM) (e.g. \citealt{finlator2008}). However, this is still a matter of some debate as, for a local sample of galaxies observed with integrated field spectroscopy, \cite{sanchez2013} do not find evidence of a negative correlation between metallicity and SFR at fixed stellar mass.

A number of recent studies have attempted to confirm the presence of this FMR at $z\gtrsim1$ with various high redshift samples. For example, using a sample of highly star-forming ($\rm>20M_{\odot}yr^{-1}$), $z\sim1$ \emph{Herschel} far-infrared selected galaxies \cite{roseboom2012} find that their galaxies are consistent with the SDSS FMR albeit with large scatter. At $z\sim1.4$ \cite{yabe2012} also find some agreement with the FMR when following up a highly star-forming ($\rm>20M_{\odot}yr^{-1}$), photometrically selected sample, but again with a significant scatter compared with $z=0.1$. Studies of small samples of gravitationally lensed systems by \cite{richard2011,wuyts2012} and \cite{belli2013}, in particular the latter, test the low mass regime of the FMR at $1.0<z<3$ and again find some agreement with the local plane. However, it is clear that a well defined sample of typical high redshift galaxies that probes down to lower SFRs ($\rm<10M_{\odot}yr^{-1}$) is required to take these comparisons further.

Investigations of the form of the FMR and its evolution are important to constrain models and simulations of galaxy evolution that invoke a feedback mechanism to regulate star-formation, thought to be an important component in our understanding of galaxy evolution (e.g.  \citealt{larson1974,white1978}). In all but the most massive galaxies, where active galactic nuclei (AGN) are often invoked to suppress star formation (e.g. \citealt{bower2006}), it is heating and winds from supernovae and massive stars that provide this regulation (e.g. \citealt{benson2003,crain2009}). Through the balance of star formation with feedback and the inflow of lower metallicity gas from the IGM, some recent models have managed to broadly reproduce the properties of the FMR at low redshift (e.g. \citealt{dave2011,dave2012,dayal2013,lilly2013}). 

Narrow-band $\rm H{\alpha}$ surveys provide a volume-selected sample (to an SFR limit) allowing for straight-forward analysis of trends with SFR or mass (e.g. \citealt{sobral2011,stott2013}). The $\rm H{\alpha}$ emission line is less affected by dust obscuration than shorter wavelength star formation tracers (e.g. UV and [OII]). Beyond $z=0.4$ $\rm H{\alpha}$ is redshifted out of the optical window, thus high redshift studies of star formation have been limited to either using the obscuration-affected short wavelength tracers or studying small samples of $\rm H{\alpha}$ emitters from conventional near-infrared spectrographs. However, in the last five years wide-field, narrow-band surveys such as the High-redshift (Z) Emission Line Survey (HiZELS, \citealt{geach2008,geach2012,garn2010a,sobral2009,sobral2010a,sobral2012,sobral2013,stott2013}) have started to provide large samples of $\rm H{\alpha}$-selected galaxies (see also the studies of \citealt{villar2008} and \citealt{ly2011}). These narrow-band surveys produce emission line information over large areas of the sky and are thus able to probe both ends of the $\rm H{\alpha}$ luminosity and stellar mass functions of star-forming galaxies, required for an unbiased analysis of the SFR density (e.g. \citealt{sobral2013}). 

HiZELS provides the large well-defined sample of star-forming galaxies at high redshift required to investigate the evolution of the FMR from the peak epoch of star formation ($z\sim1-2$). In this paper we obtain gas phase metallicities for HiZELS sources at $z=0.84$ and $z=1.47$, with the near-infrared Fiber Multi Object Spectrograph (FMOS, \citealt{fmos2010}) on the Subaru Telescope. This enables us, for the first time, to probe the mass, metallicity and SFRs of typical star-forming galaxies, in order to map out the equivalent of the FMR at this crucial epoch for galaxy evolution studies. 

The structure of this paper is as follows. In \S\ref{sec:samp} we describe the HiZELS sample, the FMOS observation data and the spectral line fitting. We then assess the SFR, metallicity and AGN content through emission line diagnostics and the stellar mass through the available broad band data (\S\ref{sec:ana}). In \S\ref{sec:results} we compare the HiZELS mass-metallicity relation to that from other studies and then test for the presence of a plane in mass, metallicity and SFR at $z\sim1-1.5$. Finally in \S\ref{sec:disc}, we discuss the implications of our findings in the context of galaxy evolution and the increase in the star formation rate density with redshift.

A $\Lambda$CDM cosmology ($\Omega_{\rm m}=0.27$, $\Omega_{\Lambda}=0.73$, $H_{0}=70$ km\,s$^{-1}$ Mpc$^{-1}$) is used throughout this work and all magnitudes are AB.   

\section{The sample \& data}
\label{sec:samp}
HiZELS \citep{geach2008,sobral2013} is a Campaign Project using the Wide Field CAMera (WFCAM, \citealt{casali2007}) on the United Kingdom Infra-Red Telescope (UKIRT) which exploits specially designed narrow-band filters in the $J$ and $H$ bands (NB$_J$ and NB$_H$), along with the H$_{2}$S1 filter in the $K$ band, to undertake panoramic, moderate depth surveys of emission line galaxies. HiZELS targets the $\rm H{\alpha}$ emission line redshifted into the near-infrared at $z = 0.84, 1.47 \rm \,and \,2.23$ using these filters. In addition, the UKIRT data are complemented by deeper narrow-band observations with Subaru Suprime-Cam NB921 imaging \citep{sobral2012,sobral2013} to obtain $\rm H{\alpha}$ emitting galaxies at $z=0.4$ and the [OII] emission from the $z=1.47$ $\rm H{\alpha}$ sample, as well as deeper WFCAM and Very Large Telescope near-infrared imaging through the H$_{2}$S1 filter in selected fields. The survey is designed to trace star-formation activity across the peak of SFR density and provide a well-defined statistical sample of star-forming galaxies at each epoch (see \citealt{best2010}).  

In this study we concentrate on the HiZELS $z = 0.84$ and $z=1.47$ $\rm H{\alpha}$ emitters in the COSMOS, Bootes and Elais-N1 fields, with the $z = 1.47$ galaxies being the primary sub-sample and the $z = 0.84$ being the secondary. These samples were observed with FMOS, a near-infrared fibre-fed multi-object spectrograph on the Subaru Telescope \citep{fmos2010}. FMOS allows for the placement of up to 400 $1.2''$ diameter fibres within a $30'$ diameter circular field of view and the light from these fibres is extracted as spectra by the two spectrographs (IRS1 and IRS2). A mirror mask is installed in the spectrographs for OH airglow suppression. The spectral coverage is from $0.9\mu \rm m$ to $1.8\mu \rm m$ which is divided into four $0.25\mu \rm m$ wide bands ($J$-Short, $J$-Long, $H$-Short and $H$-Long) with $\rm R\approx2200$, on average, in high resolution mode. 

The galaxies are drawn randomly from the HiZELS catalogues with the primary, $z=1.47$, sub-sample observed with both the FMOS $H$-Long (R=2600) filter, to measure $\rm H{\alpha}\,6563\AA\,$ and [NII]\,6583\AA\, line fluxes, and the $J$-Long (R=1900) filter, to obtain the [OIII]\,5007\AA\, and $\rm H{\beta}\,4861\AA\,$ fluxes. Those in the secondary, $z=0.84$, sub-sample are observed using the $J$-Long filter again to obtain $\rm H{\alpha}$ and [NII]. 

In total we observed six FMOS fields-of-view, integrating for one hour each in both the $J$-Long and $H$-Long wavebands on the nights of 4th and 5th May 2012 in $0.6''-0.8''$ optical seeing conditions (Proposal ID: S12A-062). These are divided into two COSMOS fields, three Bootes fields and one Elais-N1 field. In total we target \totaltarget\ $z = 0.84$ and $1.47$ $H{\alpha}$ emitters (\totaltargetJ\  and \totaltargetH\  respectively). The remainder of the FMOS fibres were used to target $z = 2.23$ galaxies for studies of [OIII] and [OII]\,3727\AA\,, non-HiZELS AGN targets, flux calibration stars or were placed on the sky to improve the removal of the atmospheric OH lines. The data for these additional targets are not discussed further in this paper.

For the observations we use the cross beam switch (CBS) technique in which two fibres are allocated to one source and the telescope is offset between two positions so that either of the two fibres observes the source and the other observes sky, for improved sky subtraction at near-infrared wavelengths (e.g. see \citealt{rodrigues2012}). This means that up to 200 sources can be observed at once. This sampling is appropriate to the target density of the HiZELS survey so the choice of CBS mode is not a limitation. The advantages of the CBS method are that targets are observed for 100\% of the time and that sky subtraction is less affected by the temporal and spatial variation of sky brightness.

\subsection{Data reduction}

The data are reduced with the Subaru FMOS reduction pipeline, FMOS Image-Based REduction Package ({\sc fibre-pac}, \citealt{fibrepac2012}). {\sc fibre-pac} is a combination of {\sc iraf} tasks and {\sc c} programs using the {\sc cfitsio} library \citep{pence1999}. 

\begin{figure*}
   \centering
\includegraphics[scale=0.275, trim=0 90 20 40, clip=true]{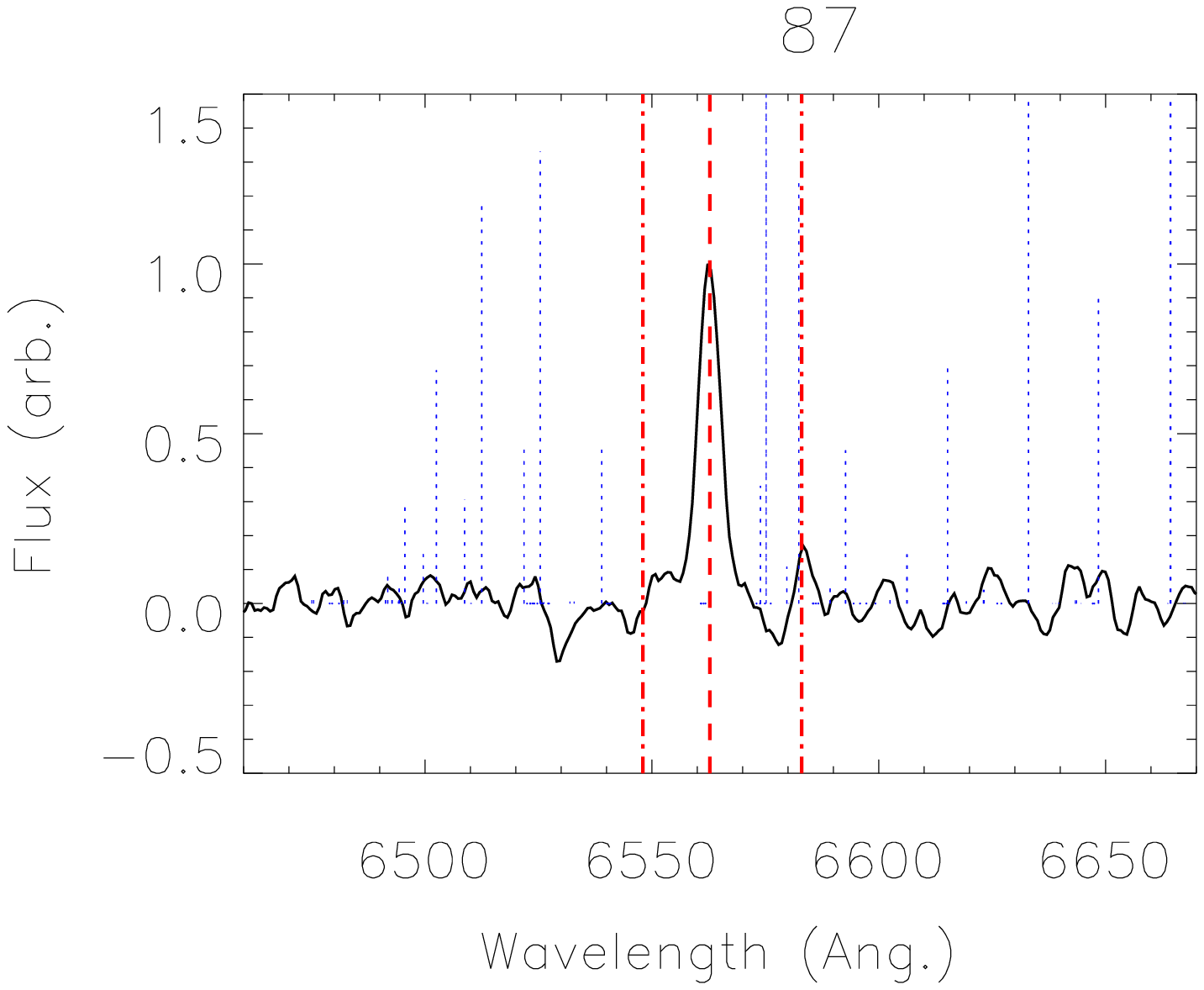}
\includegraphics[scale=0.275, trim=105 90 20 40, clip=true]{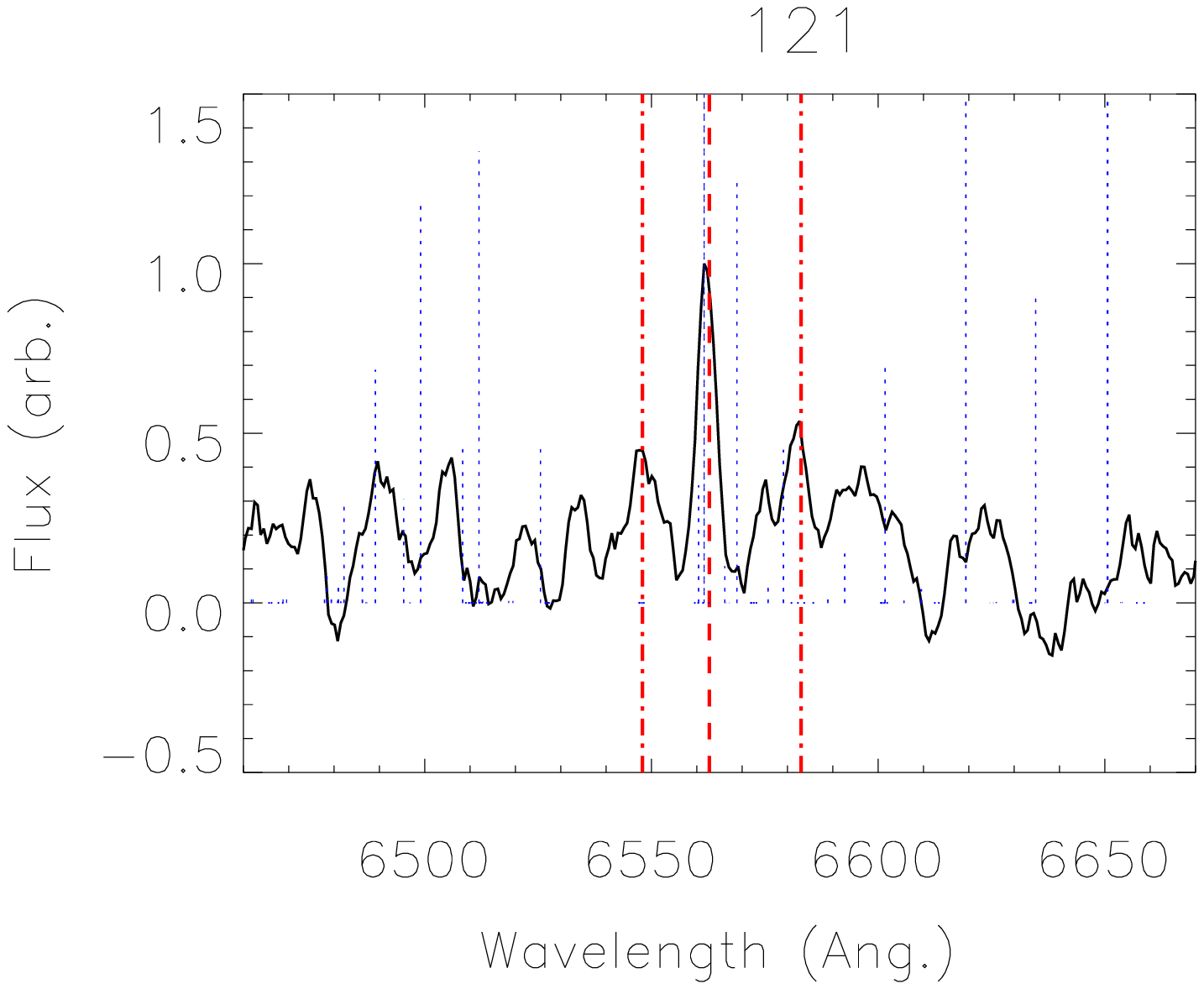}
\includegraphics[scale=0.275, trim=105 90 20 40, clip=true]{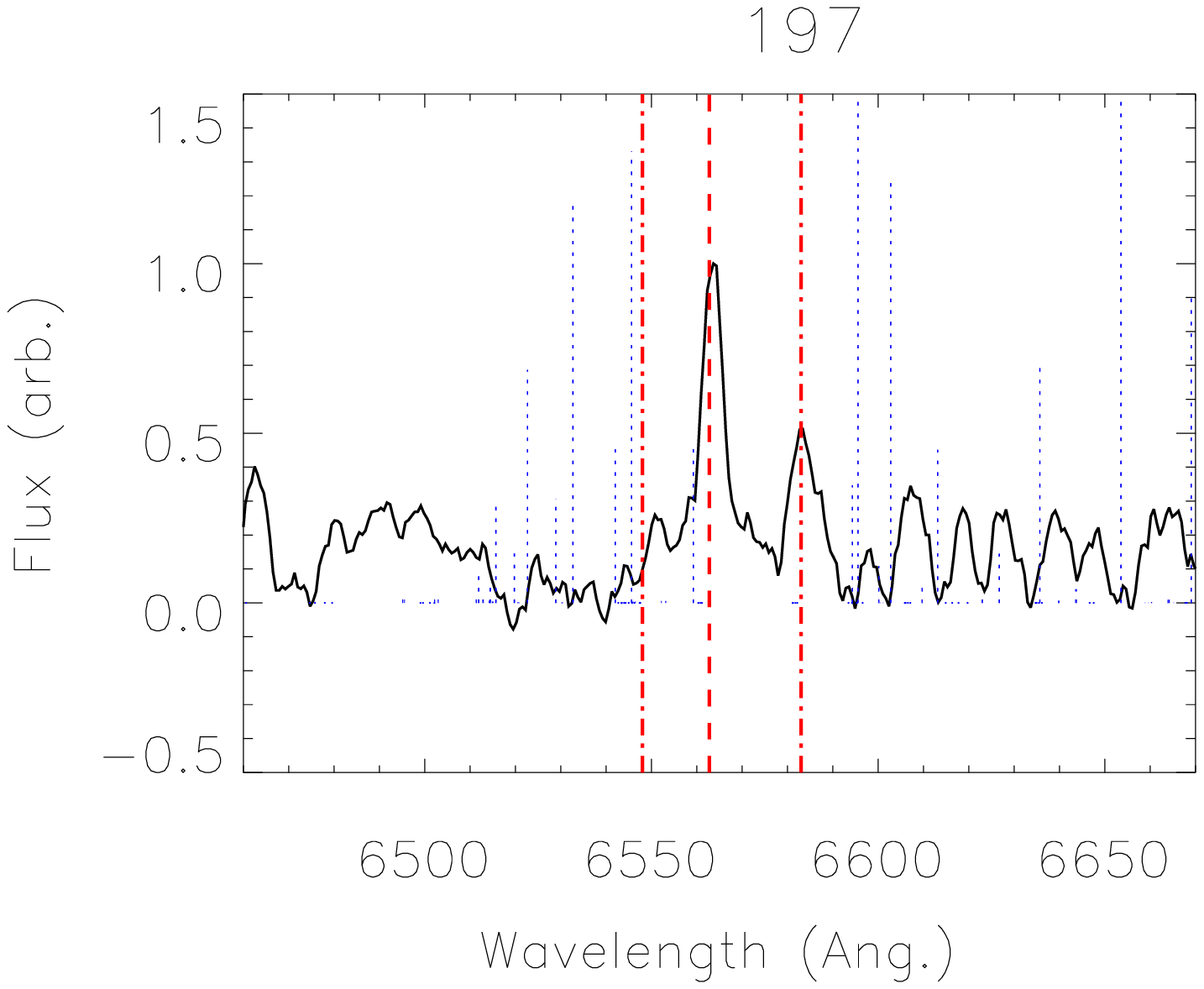}
\includegraphics[scale=0.275, trim=105 90 20 40, clip=true]{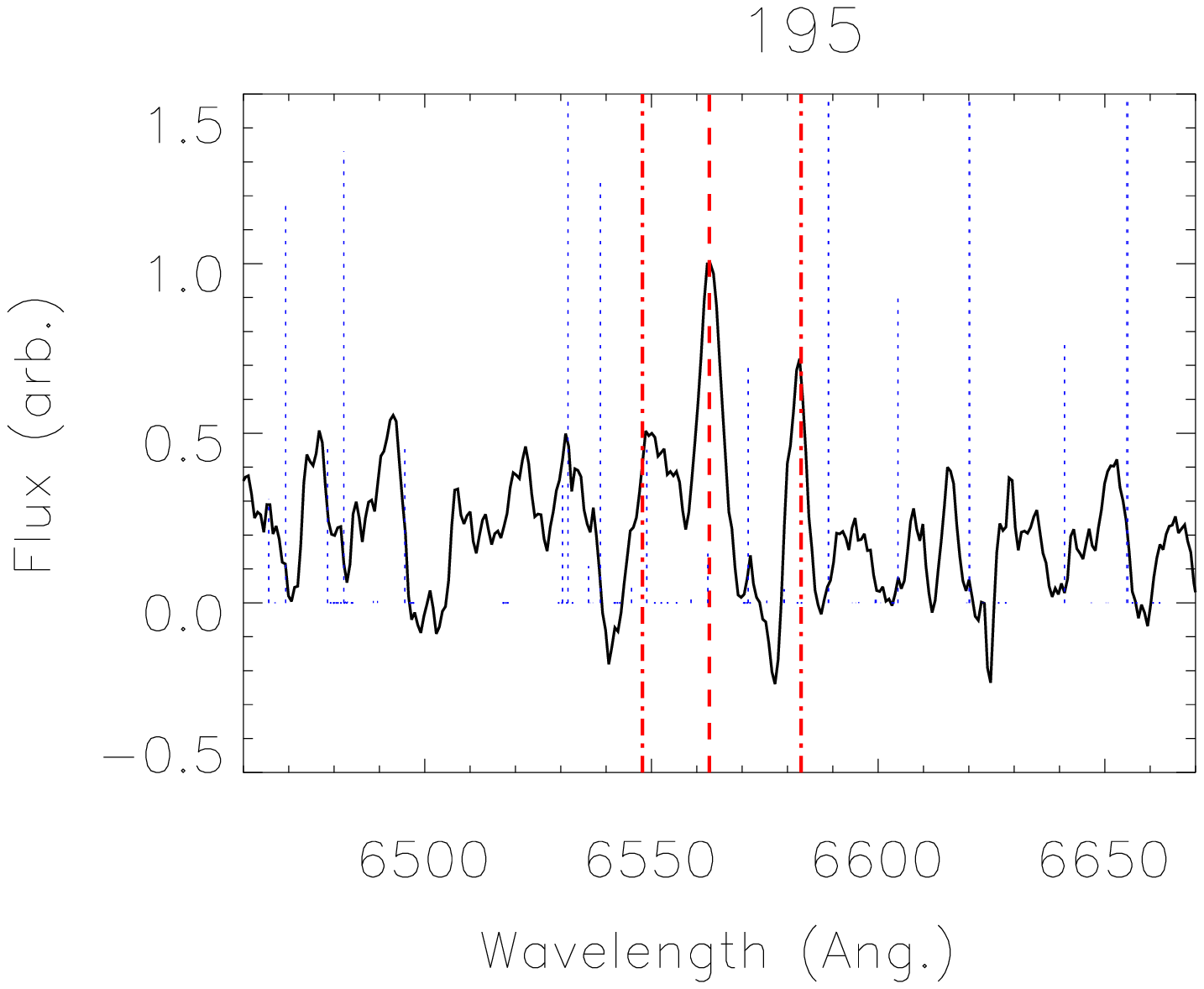}
\includegraphics[scale=0.275, trim=0 90 20 40, clip=true]{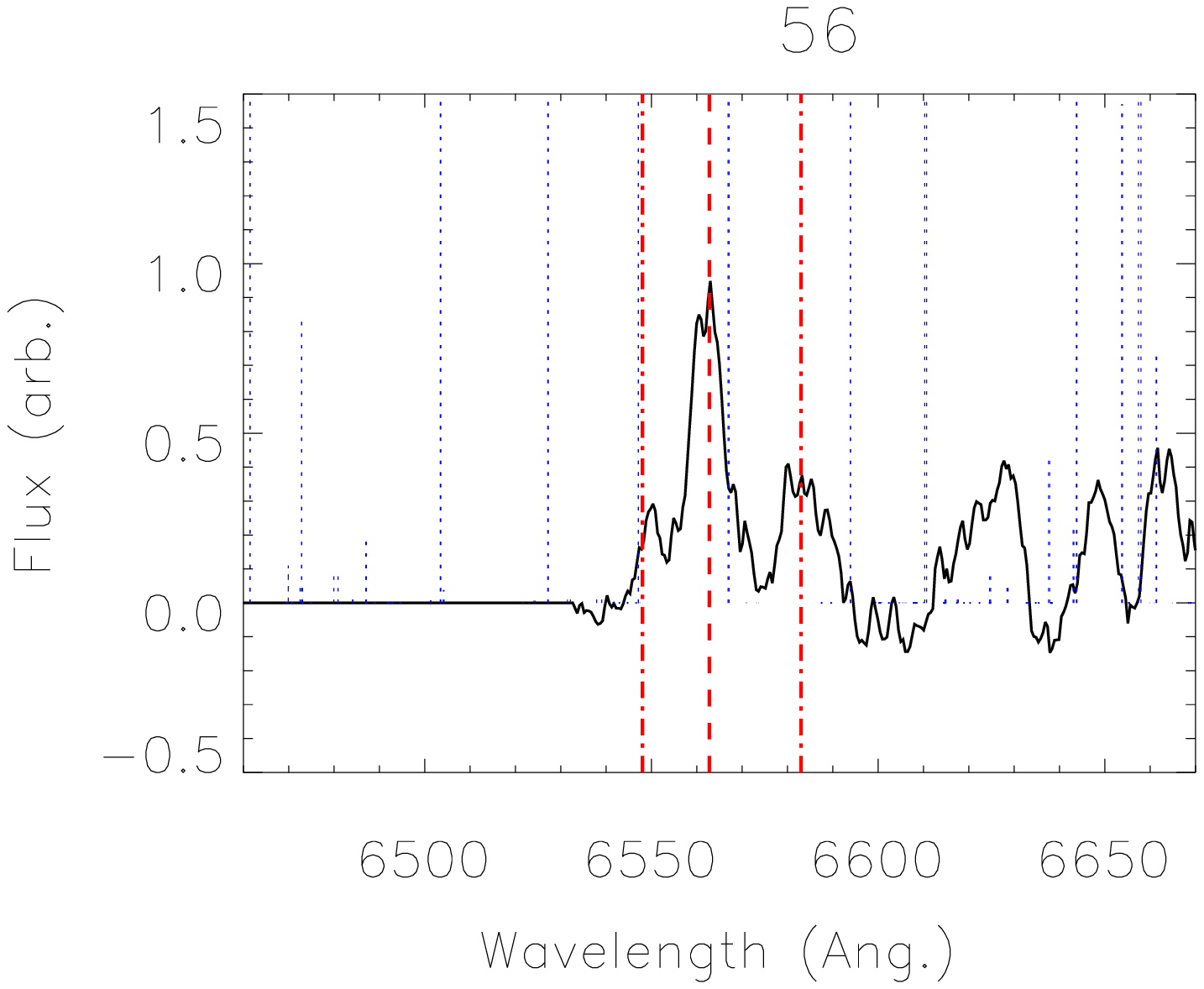}
\includegraphics[scale=0.275, trim=105 90 20 40, clip=true]{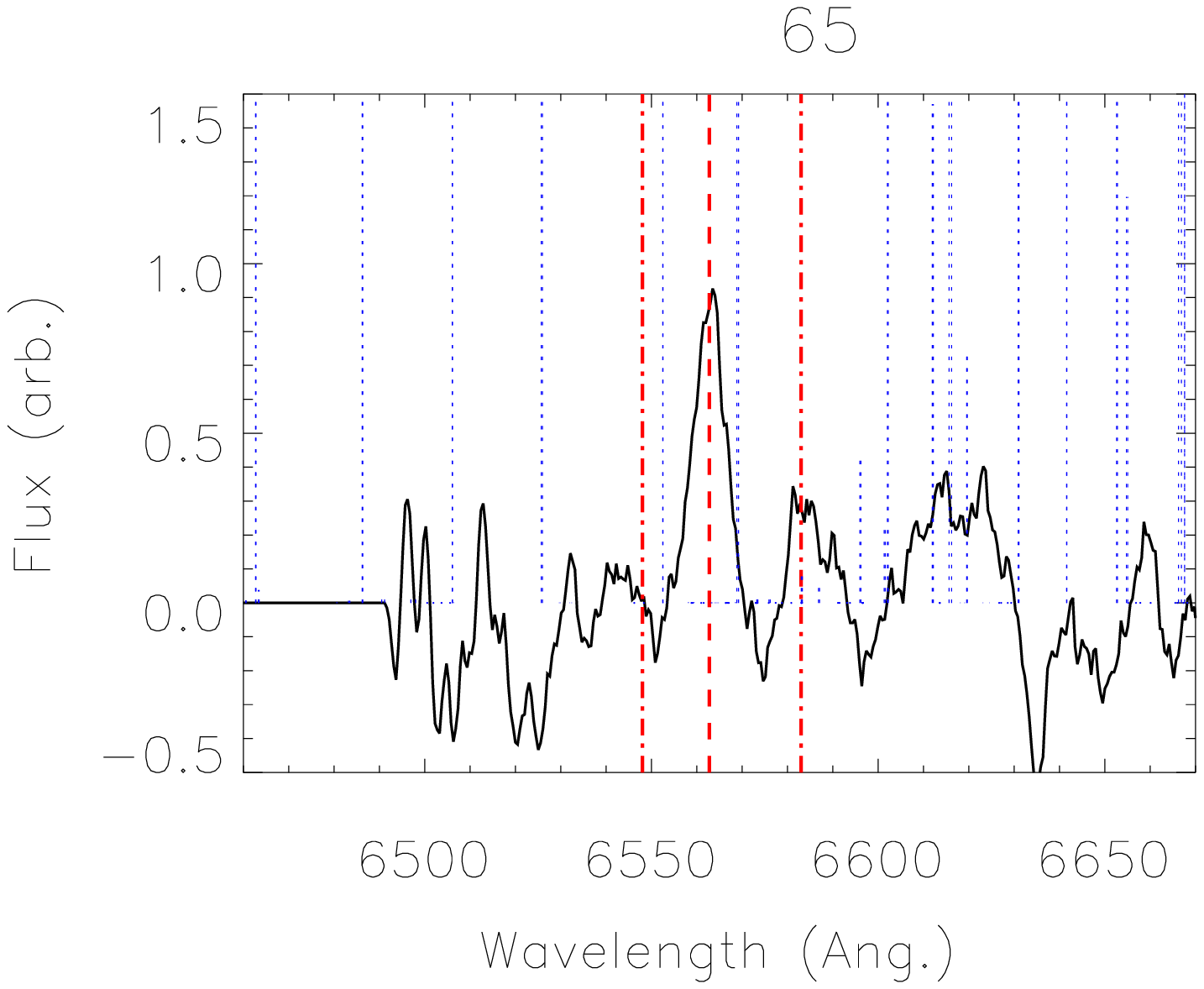}
\includegraphics[scale=0.275, trim=105 90 20 40, clip=true]{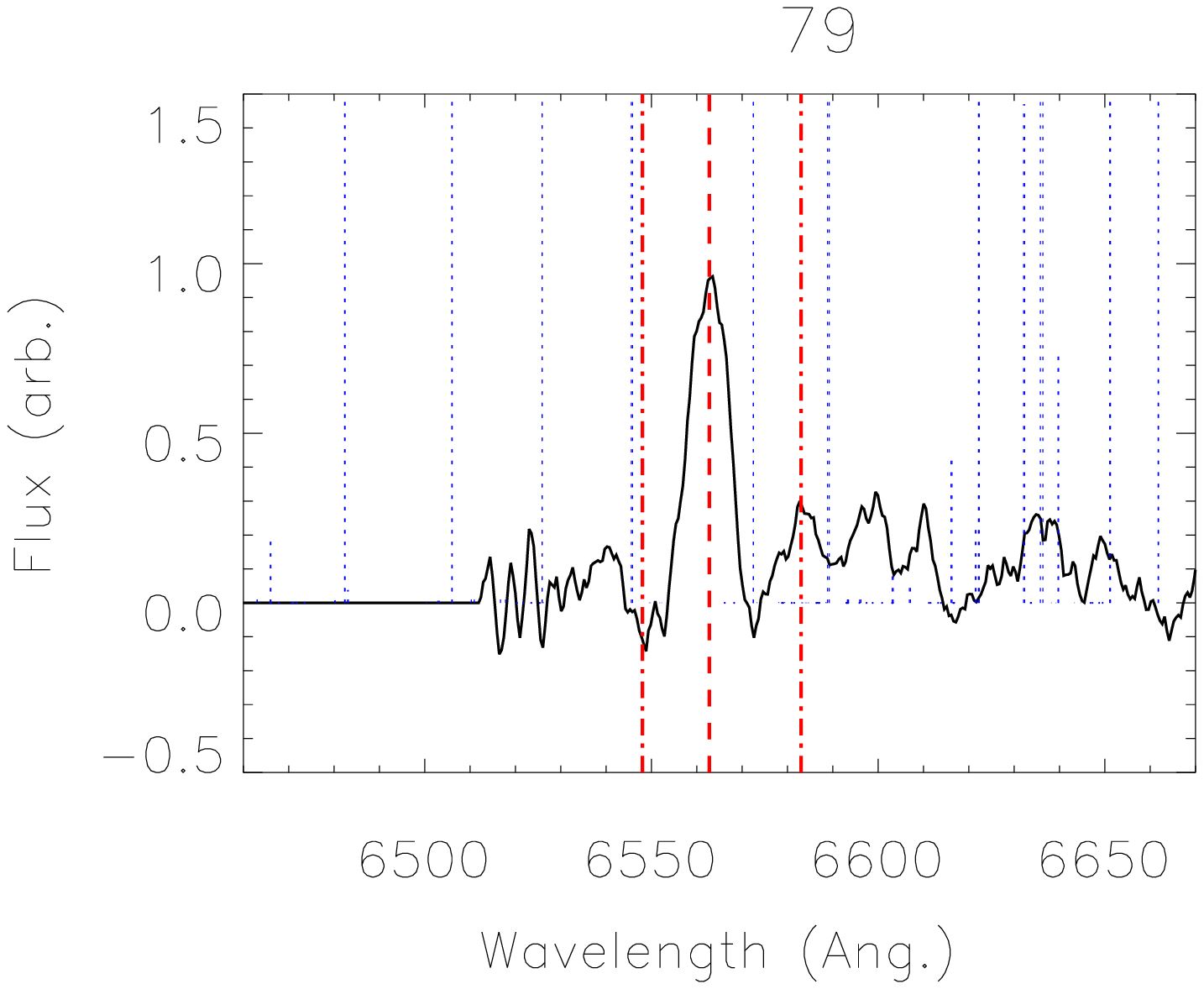}
\includegraphics[scale=0.275, trim=105 90 20 40, clip=true]{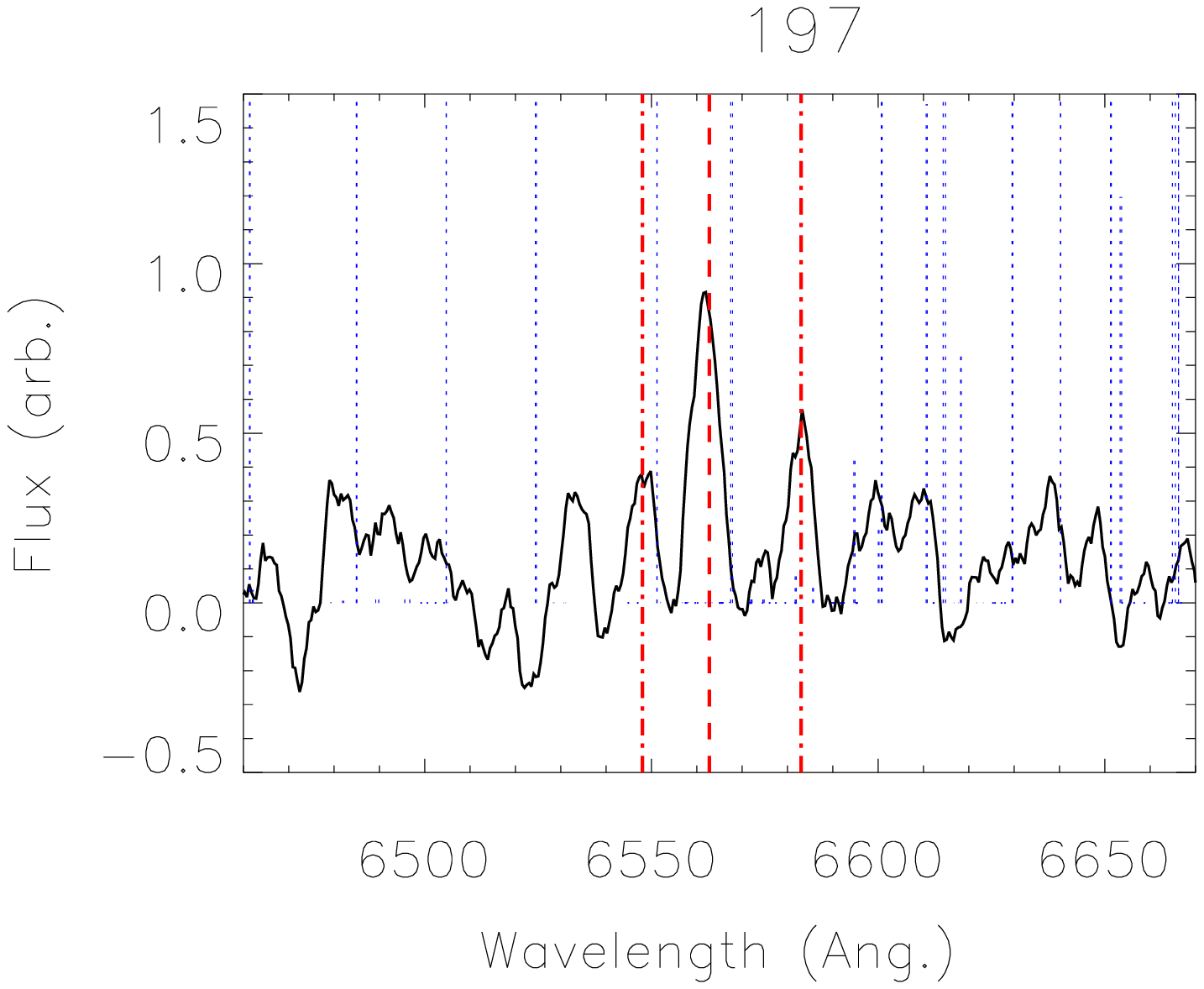}
\includegraphics[scale=0.275, trim=0 0 20 40, clip=true]{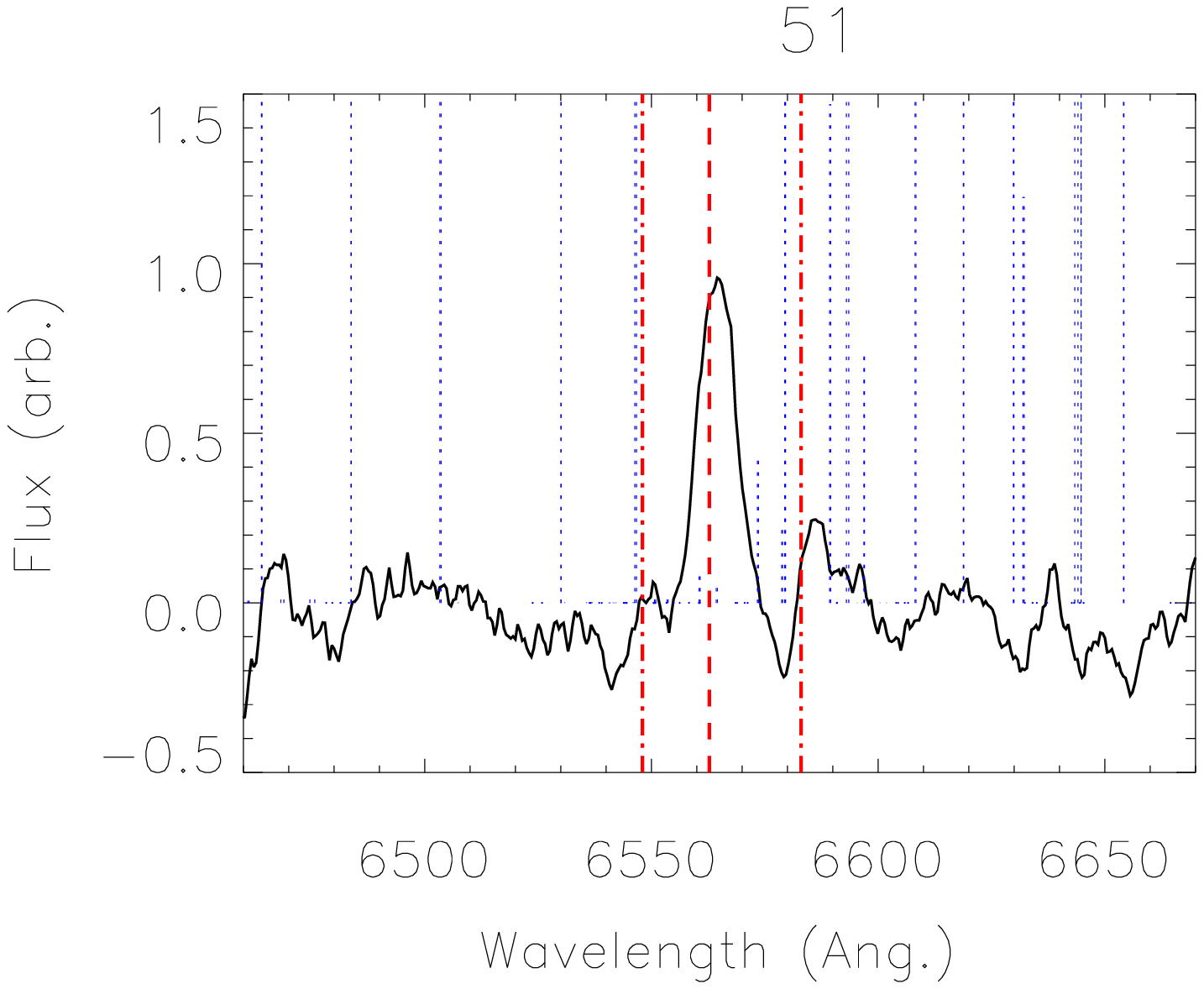}
\includegraphics[scale=0.275, trim=105 0 20 40, clip=true]{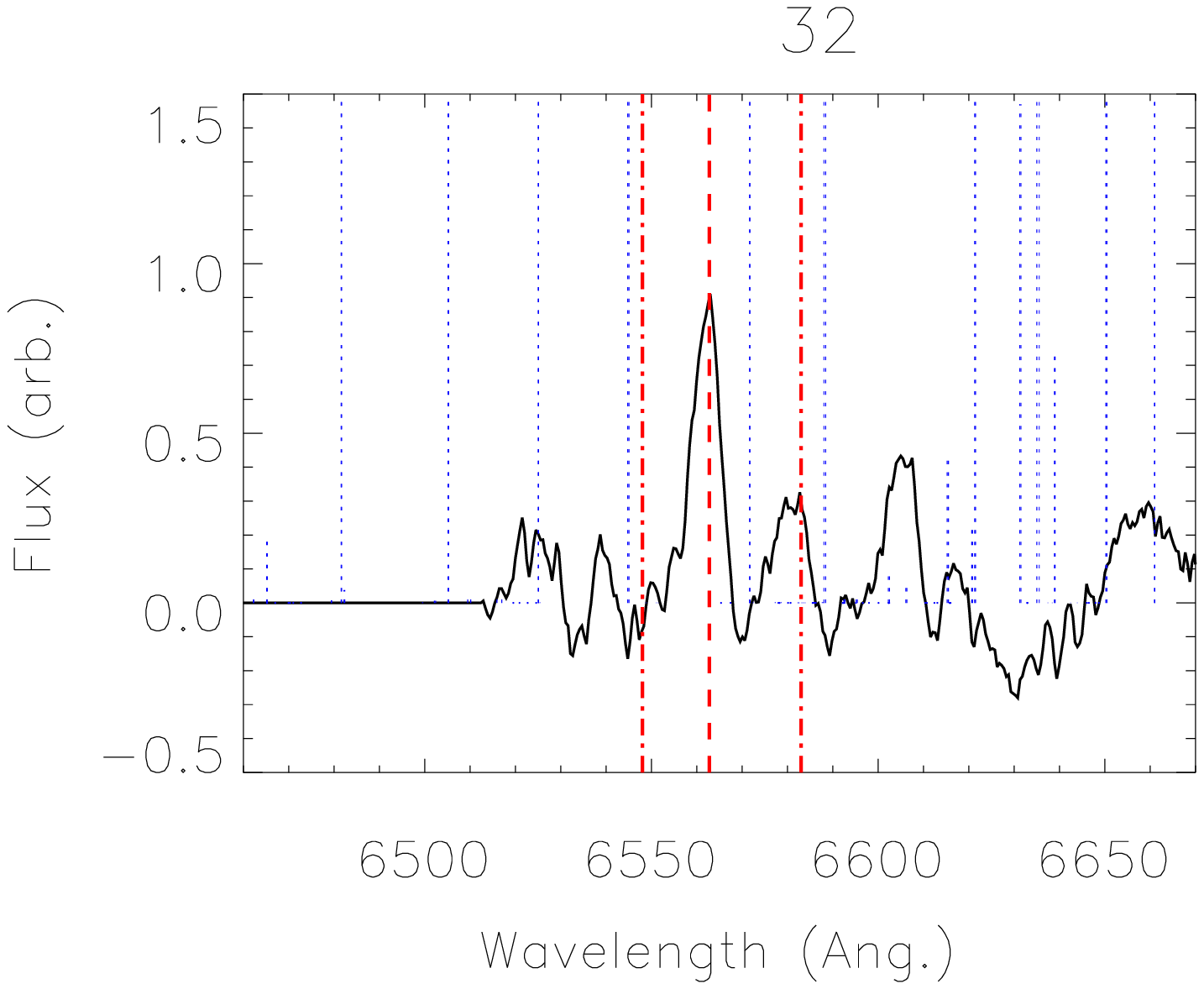}
\includegraphics[scale=0.275, trim=105 0 20 40, clip=true]{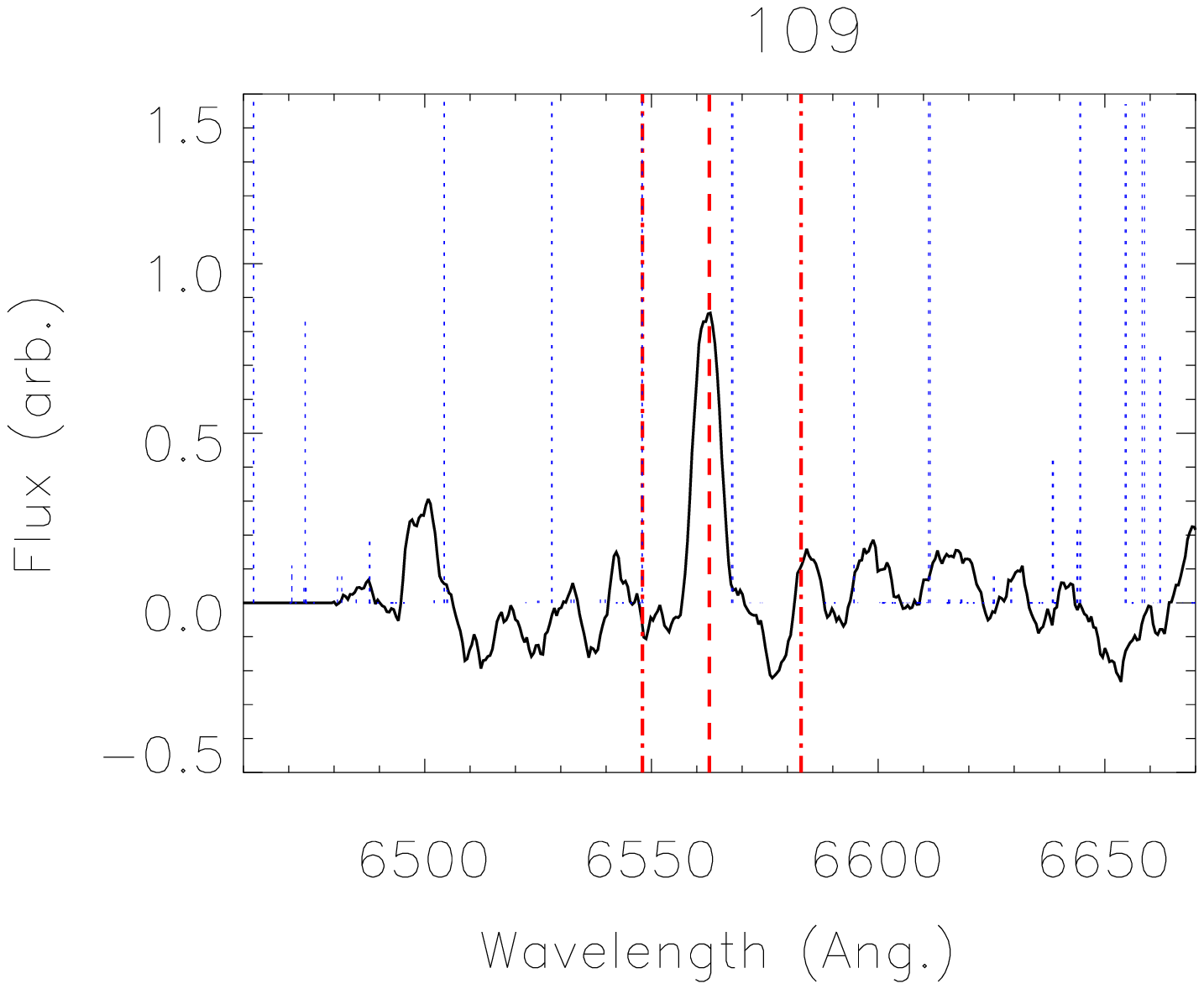}
\includegraphics[scale=0.275, trim=105 0 20 40, clip=true]{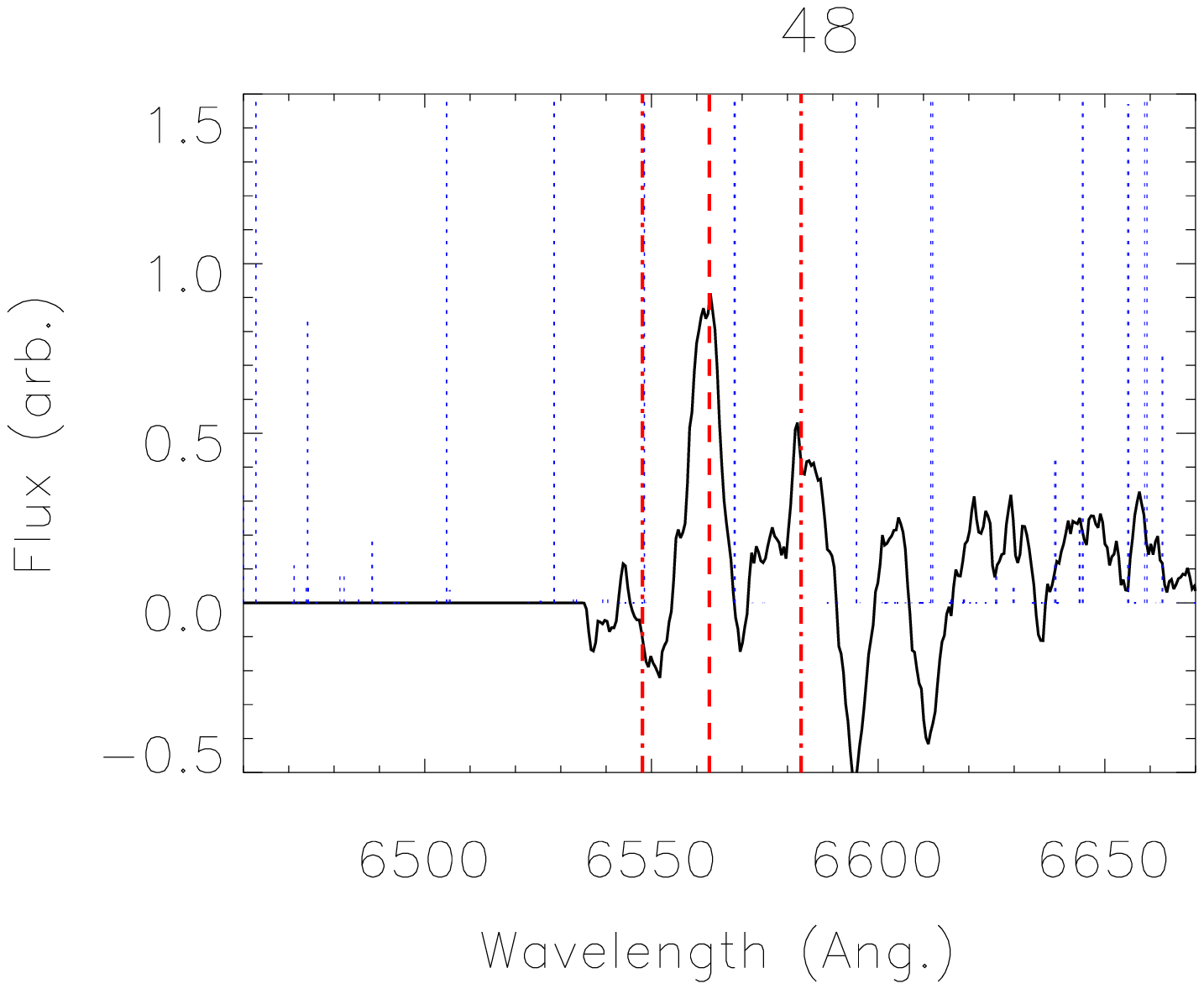}
\caption[]{A selection of 12 spectra taken from the sample of \totalHaNIItwosig\ HiZELS-FMOS galaxies for which we recover both $\rm H{\alpha}$ and [NII] emission lines with, $\rm S/N>5$ and $2$ respectively (see \S\ref{sec:stat}). The wavelength is given in the rest-frame, the flux is normalised to the peak flux of the $\rm H{\alpha}$ emission line and a smoothing of 1.5 FMOS resolution elements has been applied (9\AA). The vertical dotted blue lines show the location of the sky lines, in the {\it observed frame}, from the \cite{rousselot2000} catalogue and we have removed the FMOS OH suppression mask where it affects the data by dividing by the high noise values at their locations. The wavelengths of the $\rm H\alpha$ and [NII] (6548\AA\, and 6583\AA) spectral lines are marked by the dashed and dot-dashed red lines respectively.}
   \label{fig:spec}
\end{figure*}

The raw data consist of science frames, dome-flats and Th-Ar spectral calibration arcs. The details of the reduction process can be found in the {\sc fibre-pac} paper \citep{fibrepac2012} but we outline the key steps here. In the initial preparation the data are first flat fielded and bad pixels removed. Corrections are applied to remove spatial and spectral distortions present and then wavelength calibration is performed. An initial background subtraction is then achieved using the ABAB nodding pattern of the telescope to perform an A-B sky subtraction. Further bad pixel, detector cross talk, bias difference, distortion, residual background and sky corrections are then applied. The 2-d spectra are then combined, which in CBS mode means inverting and adding the negative B spectra to the A spectra. The final step is an initial flux calibration (see \S\ref{sec:fit}) using the spectra from a number of fibres which were set to observe stars with known near-infrared spectral energy distributions (i.e. those with The Two Micron All Sky Survey [2MASS] photometry, \citealt{2mass}) in combination with template stellar spectra from \cite{rayner2009} and accounting for the effects of atmospheric absorption. The fully reduced 1-d spectra corresponding to each fibre pair are then extracted from the 2-d frame. A selection of individual, example spectra used for the main sample in this paper, covering the full range in signal-to-noise, are provided in Fig. \ref{fig:spec}.

\subsection{Spectral line fitting}
\label{sec:fit}
The advantage of the HiZELS survey is that the redshift of the galaxies in our sample is known to $\delta z\sim\pm0.015$ and therefore it is not necessary to search the entire wavelength range for emission lines. We take the 1-d spectra, smoothed over 6\AA\, (corresponding to 5 spectral pixels, appropriate to the spectral resolution of $\sim2200$), and identify the emission lines that fall in the wavelength range of the HiZELS narrow-band filters. These emission lines are fit with a single Gaussian profile in order to extract their total flux. We extract both the $\rm H{\alpha}$\,6563\AA\, and [NII]\,6583\AA\, emission lines for the $z=1.47$ and $z=0.84$ samples from the $H$-Long and $J$-Long observations respectively with these lines easily resolved from each other in high resolution mode. For the $z=1.47$ galaxies we also fit to any spectral lines present at the wavelengths corresponding to $\rm H{\beta}$\,4861\AA\, and [OIII]\,5007\AA\, line emission within the appropriate $J$-Long observations.

As with all near-infrared observations these spectra are affected by OH sky lines that were not fully removed in the data reduction pipeline. A further important factor with FMOS is that there is also a mirror mask to suppress the strongest OH airglow lines. This means that in the regions where there is a strong sky line present the flux has been completely removed. Because of the sky lines and the OH suppression it is difficult to extract real spectral lines that are in the vicinity of the sky affected regions, with some potentially being completely obscured in regions where the OH suppression is most aggressive (see also \citealt{yabe2012}). We flag galaxies with spectral lines that are strongly affected by the OH suppression, i.e. with central wavelengths that fall within one FMOS resolution element (6\AA) of a line in the OH suppression mask, these galaxies are then removed from any analysis which requires that specific line.

After obtaining initial fluxes from the FMOS spectra we then apply a further flux calibration based on the known $\rm H{\alpha}$ HiZELS narrow-band fluxes of the galaxies. As well as providing a more accurate flux calibration, this accounts for the  $1.2''$ diameter FMOS fibres which potentially means that not all of the $\rm H{\alpha}$ emission for a given galaxy detected in the HiZELS survey will be observed by FMOS. For example, the $\rm H{\alpha}$ emission from the galaxy may extend outside of this diameter, as from \cite{stott2013} we know that the average broad-band half-light radii of the HiZELS galaxies at these redshifts is $\sim3-4\rm kpc$ or $\sim0.5''$. Another related possibility is that the fibres are not centred exactly on the peak of the $\rm H{\alpha}$ emission. We use an individual aperture correction for each galaxy, which is the ratio of the HiZELS narrow-band flux to the $\rm H{\alpha}$ + [NII] flux measured from the FMOS spectroscopy. The median value of this aperture correction, for the full sample of \totalHaNIItwosig\ galaxies with detected $\rm H{\alpha}$ and [NII] used throughout this paper (see \S\ref{sec:stat}), is $2.6\pm0.2$ (in agreement with that found by \citealt{roseboom2012,yabe2012,matsuda2011}). 

We calculate the signal-to-noise ratio (S/N) of the $\rm H\alpha$ measurements from the noise image produced by the data reduction pipeline and the variance of the background of the calibrated spectra. We only consider $\rm H\alpha$ lines with a S/N$>5$, which corresponds to a flux $\sim4\times10^{-17} \rm erg \,cm^{-2} s^{-1}$ for both the $J$-Long and $H$-Long set ups, in agreement with both the values quoted for 1 hour of integration in the FMOS documentation and with the limiting flux of the input HiZELS narrow-band sample. The corresponding S/N=5 SFR limits for $z=0.84$ and $z=1.47$ are therefore $\sim1 \,\rm M_{\odot}yr^{-1}$ and $\sim5\, \rm M_{\odot}yr^{-1}$ respectively, assuming a \cite{kennicutt1998} star formation law, corrected to a \cite{chabrier2003} initial mass function (IMF) and a dust extinction at the $\rm H\alpha$ wavelength, $A_{H{\alpha}}=1$\,mag (see \S\ref{sec:ana}). We note here that recent, similar studies with FMOS have only probed down to $\sim1\times10^{-16} \rm erg \,cm^{-2} s^{-1}$ ($\gtrsim20 \rm M_{\odot}yr^{-1}$ at $z\sim1.5$) due to the use of the FMOS low-resolution mode (i.e. \citealt{roseboom2012,yabe2012}) rather than the more sensitive high-resolution mode utilised here.

An additional consideration when studying $\rm H\alpha$ and $\rm H\beta$ emission lines is that they are superimposed on the stellar Balmer absorption line features. As we do not detect significant continuum in either our individual or stacked galaxy spectra we cannot assess this directly. However, following \cite{yabe2012} we note that both \cite{savaglio2005} and \cite{zahid2011} find this effect to be small, typically requiring a correction of  $1-4$\AA\, to the equivalent width (EW) of the emission line. HiZELS has a rest-frame EW lower limit of 25\AA\, and a median rest-frame EW of $\sim100$\AA\, and thus the effect of the Balmer absorption is negligible by comparison. We therefore choose not to include a correction for this effect.

\subsection{Sample statistics \& accounting for non-detections}
\label{sec:stat}
\begin{figure}
   \centering
\includegraphics[scale=0.5, trim=0 55 0 0, clip=true]{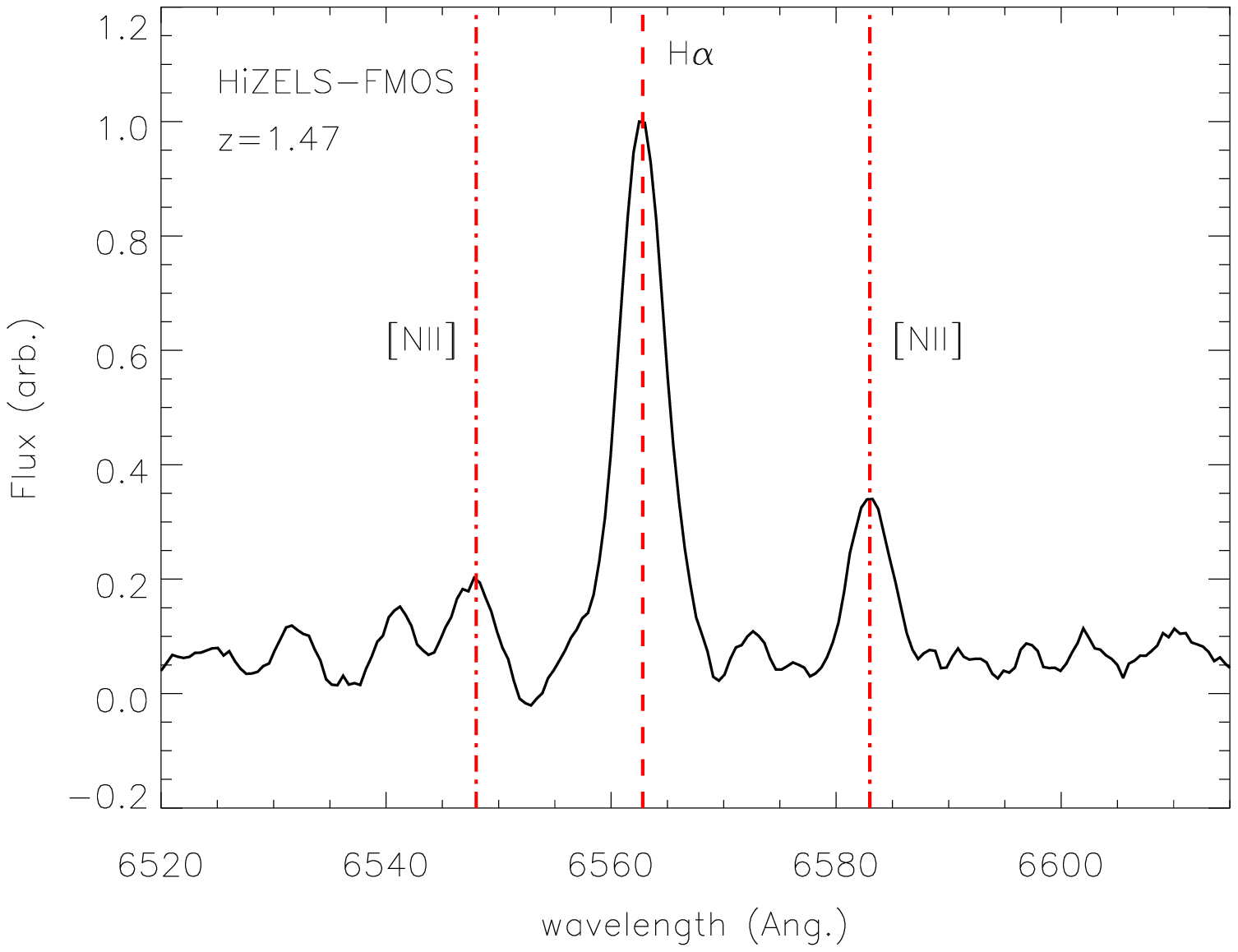} \includegraphics[scale=0.5, trim=0 0 0 27, clip=true]{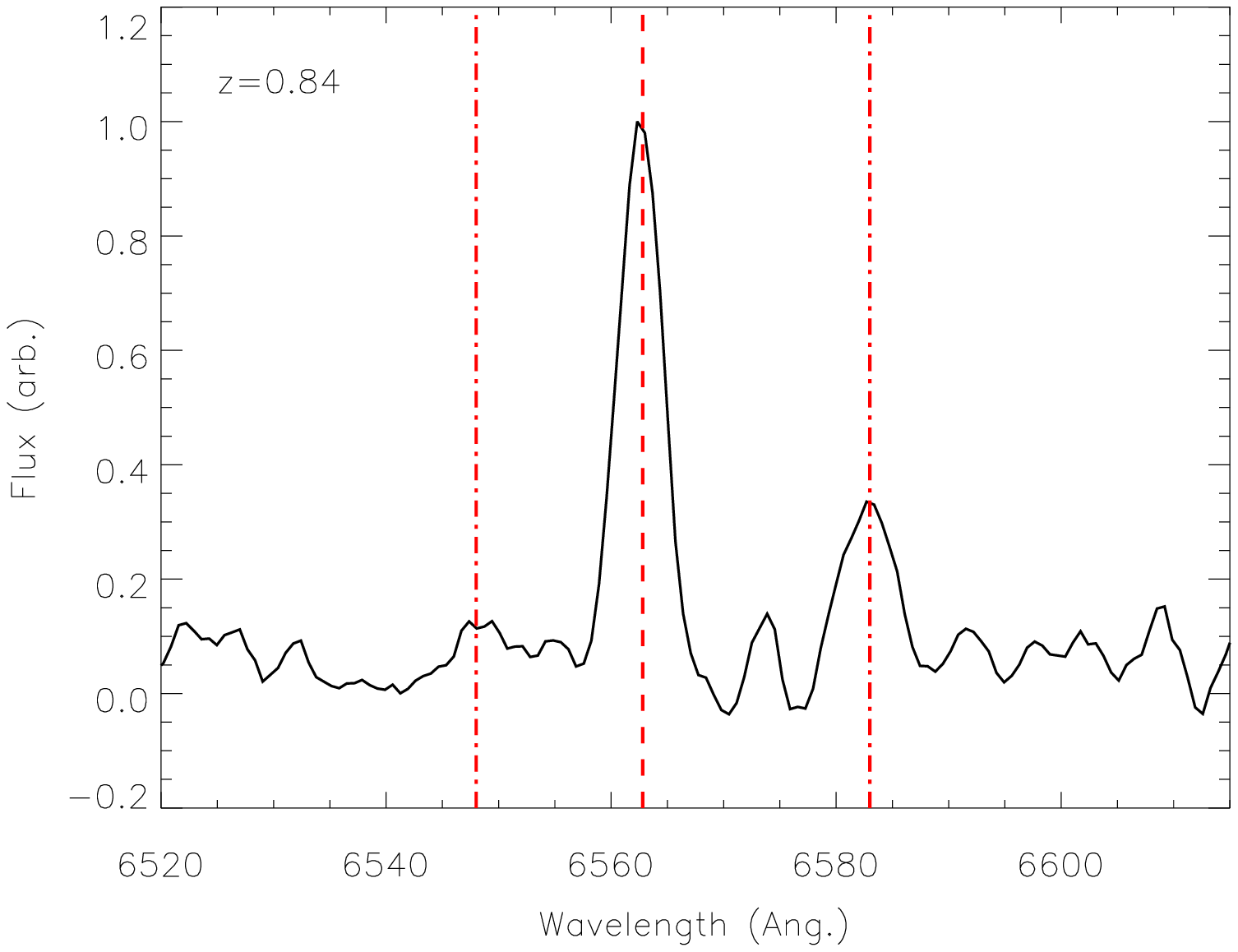} 

\caption[]{A rest-frame stack of the spectra of the $z=1.47$ and $z=0.84$ $\rm H\alpha$ emitting galaxies with the wavelength of the $\rm H\alpha$ and [NII] (6548\AA\, and 6583\AA) spectral lines marked by the dashed and dot-dashed red lines. The stacked ratios of $\rm[NII]/ H\alpha$ are $0.36\pm0.03$ and $0.34\pm0.02$ for $z=0.84$ and $z=1.47$ respectively, corresponding to the metallicities $12+\log\rm(O/H)=8.65\pm0.02$ and $8.63\pm0.02$ (assuming the conversion of \citealt{Pettini2004}), consistent with the solar value of $8.66\pm0.05$ \citep{asplund2004}.}
   \label{fig:HaNII}
\end{figure}

Of the \totaltarget\ $z = 0.84$ and $1.47$ HiZELS $\rm H{\alpha}$ emitters targeted, we recover $\rm H{\alpha}$ emission above S/N=5 for \totalHafivesig\ ($\sim50\%$) and $\rm H{\alpha}$ + [NII] for \totalHaNIItwosig\ ($\sim33\%$) of these (with a S/N=2 detection threshold for [NII]). Of the \totalHafivesigH\ for which we recover $z=1.47$ $\rm H{\alpha}$ emission in the $H$-Long set-up, we detect [OIII] in \totalHaOIIItwosig\ and $\rm H\beta$ in \totalHaHbtwosig\ (with a S/N=2 detection threshold for both) with the $J$-Long. We show some example spectra around $\rm H\alpha$ and [NII] in Fig. \ref{fig:spec}. In Fig. \ref{fig:HaNII} we present the rest frame wavelength median stacks normalised to their peak flux density for the $z = 0.84$ and $1.47$ $\rm H{\alpha}$ emitters which clearly also shows the [NII]6583\AA \,line. However, in the entire $z=1.47$ sample there are only \totalBPT\ galaxies with a measurement of all four lines ($\rm H{\alpha}$, $\rm H{\beta}$, [NII] and [OIII]) above the stated S/N thresholds. The main reason for this is that the $\rm H{\beta}$ line, and to a lesser extent the [NII] and [OIII] lines, are in wavelength ranges that are strongly affected OH lines (see also \citealt{yabe2012}).
 
It is important to assess the reasons for the \totalmissingHa\ missing HiZELS $\rm H{\alpha}$ emitters in order to characterise the HiZELS narrow-band sample. The minimum flux of the HiZELS narrow-band sample is $\sim4\times10^{-17} \rm erg \,cm^{-2} s^{-1}$, the same as our FMOS S/N=5 detection threshold, so perhaps this is due to extended flux being missed by the FMOS fibres, a fibre misalignment or the lines being coincident with an OH sky line (see \S\ref{sec:fit}), some of the sources may have fallen below this threshold or have been masked from the analysis. The possibility that we are missing galaxies due to low observed flux is confirmed when we consider how the fraction of non-detections varies with HiZELS narrow-band flux. We detect $\rm H{\alpha}$ in $\sim85\%$ of the galaxies with a narrow-band flux  greater than $\sim4\times10^{-16} \rm erg \,cm^{-2} s^{-1}$ but only recover $\sim45\%$ of those near the detection limit of $\sim4\times10^{-17} \rm erg \,cm^{-2} s^{-1}$.  If we now consider all `detections' of potential $\rm H{\alpha}$ (i.e. where a line was fitted to a spectral feature in the correct wavelength range, not associated with an OH sky line) above a S/N=2 threshold this accounts for \missingHaduetolowsig\ of the missing galaxies ($\sim50\%$). We therefore conclude that half of the $\rm H{\alpha}$ non-detections are due to galaxies with a low observed flux compared to their narrow-band flux (see \S\ref{sec:fit}). The remaining \missingleft\ galaxies that are undetected in $\rm H{\alpha}$ account for just $\sim20\%$ of the total galaxies targeted. 

To measure the effect the OH sky lines have on $\rm H{\alpha}$ detection, we assess the percentage of the wavelength coverage of the narrow-band filters affected by strong OH lines and the FMOS OH suppression. To quantify this we use the OH sky line catalogue of \cite{rousselot2000} and the wavelengths of the FMOS OH suppression. From our experience with the FMOS spectra the strong, OH suppressed, sky lines and those with a relative intensity $>20$ in the \cite{rousselot2000} catalogue have a destructive affect on the spectra. Taken in combination, the percentage of the NB$_J$ and NB$_H$ filters affected by sky lines with a relative intensity $>20$ \citep{rousselot2000}, assuming the FMOS resolution of 6\AA, is $\sim20\%$. As this agrees with the percentage of unaccounted-for non-detections of $\rm H{\alpha}$ then this is most likely the reason for their absence from the HiZELS-FMOS sample.

The galaxies with detected $\rm H{\alpha}$ but without [NII] emission will have a greater impact on our study of the metallicity, which is determined from the flux ratio [NII]/$\rm H{\alpha}$, with unaccounted-for non-detections potentially acting to bias our sample to higher metallicities. The two major reasons for the non-detection of [NII] are that the line itself is faint or that it is coincident with an OH sky line, which are removed from our analysis (see \S\ref{sec:fit}). To test the relative numbers of these we compare the [NII] redshifts of all of the \totalHafivesig\ galaxies for which we detect $\rm H{\alpha}$ above a S/N=5 to the sky line catalogue of \cite{rousselot2000} and the FMOS OH suppression of the strong sky lines. To quantify the effect of the OH sky lines we calculate the number of non-detected [NII] lines that would fall within 9\AA\,($1.5\times$ the spectral resolution of FMOS at these wavelengths) of an OH line, with a relative intensity $>20$ as defined in the \cite{rousselot2000} catalogue. The median spectral separation to a sky line from the [NII] wavelength of a galaxy which is detected in $\rm H{\alpha}$ but undetected in [NII] is 6\AA\, (i.e. the resolution of FMOS), whereas for those with detected [NII] this spectral separation is 16\AA. This clearly shows a strong link between the presence of atmospheric OH emission and the non-detection of [NII]. In total 70\% of the [NII] non-detections lie within 9\AA\, of an atmospheric OH line, which we therefore conclude is the reason for their non-detection. There may also be some contamination from non-$\rm H{\alpha}$ emission line galaxies for which we would never detect [NII] although the majority of these should be removed by the sample cleaning processes described in \cite{sobral2013}. In the \S\ref{sec:sfrmassmet}, \S\ref{sec:mmr} and \S\ref{sec:fmr} analyses we include the small effect which the remaining 30\% (44) of the [NII] non-detections have on the [NII]/$\rm H{\alpha}$ flux ratio and metallicity. 

We also quantify the missing [OIII] and $\rm H\beta$ emitters in the same way. As with [NII], 70\% of the missing [OIII] and $\rm H\beta$ emitters lie within 9\AA\, of an atmospheric OH line which we again attribute as the reason for the majority of the non-detections of these lines. The median spectral separation to a sky line for a non-detected [OIII]  or $\rm H\beta$ source is again 6\AA, with the detected sources having a median spectral separation of 14\AA\, and 24\AA\, respectively. 

\section{Analysis}
\label{sec:ana}

\subsection{Dust extinction \& AGN fraction}
\label{sec:dustagn}
For the \totalHaHbtwosig\ galaxies with detections of both $\rm H\alpha$ and $\rm H{\beta}$, plus the upper limits from the non-detections of $\rm H{\beta}$ (defined by the detection threshold, S/N=2) for the remaining $\rm H\alpha$ detections that are not affected by sky lines, it is possible to estimate the average dust extinction via the Balmer decrement. The median Balmer decrement, including the upper limits, is found to be $4.2\pm0.6$. This is converted to an average dust extinction at the $V$ band wavelength, $A_V=1.3\pm0.2$, assuming an intrinsic $\rm H\alpha/H{\beta}$ ratio of 2.86 from Case B recombination \citep{osterbrock1989} and a \cite{calzetti2000} reddening curve using the same technique as \cite{dominguez2013} and \cite{momcheva2013} which accounts for the difference between the extinction derived from the nebula emission and that from the stellar continuum. This translates to an $\rm H\alpha$ extinction, $A_{\rm H\alpha}=1.1\pm0.2$. 

We also estimate the dust extinction by performing a mean average of the Balmer decrement over the galaxies with detected $H{\beta}$ and a mean stacked spectra at the $H{\beta}$ wavelength of the non-detections. This method yields  $A_V=1.4\pm0.3$ and $A_{\rm H\alpha}=1.1\pm0.3$, in agreement with the median of the detections and upper limits above. These results are entirely consistent with the average $A_{\rm H\alpha}=1$ measured by \cite{garn2010a} for the HiZELS sample and used in both \cite{sobral2013} and \cite{stott2013} and which we now adopt for the remainder of this paper for consistency with those studies. Unfortunately, the number of combined $\rm H\alpha$ and $\rm H{\beta}$ detections in the HiZELS-FMOS sample is too small to test correlations between $A_{\rm H\alpha}$ and SFR or stellar mass.

\begin{figure}
   \centering
\includegraphics[scale=0.5]{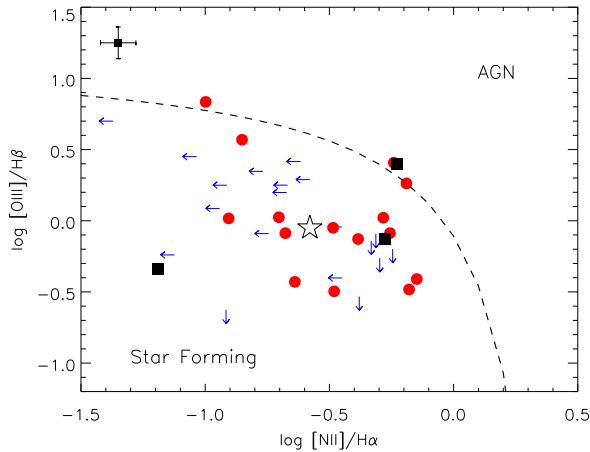} 

\caption[]{The BPT diagram \citep{baldwin1981} for the $z=1.47$ HiZELS-FMOS sources. The dashed line is the demarcation between starbursts and AGN from \cite{kew2001}. The filled black squares represent galaxies with all four emission lines i.e. $\rm H{\alpha}$, $\rm H{\beta}$, [NII] and [OIII], the upper limits (arrows) represent those with three lines that are missing either [OIII] or [NII]. The filled red circles are those missing $\rm H{\beta}$ for which we have estimated $\rm H{\beta}$ through the $\rm H{\alpha}$ flux, assuming $A_{\rm H\alpha}=1$ (see \S\ref{sec:dustagn}). The typical error is shown in the top left corner of the plot. This demonstrates that the fraction of HiZELS galaxies that occupy the same region of the BPT diagram as AGN is $\sim10\%$, in agreement with other studies (e.g. \citealt{garn2010a}). The median values of [NII]/$\rm H{\alpha}$ and [OIII]/$\rm H{\beta}$ are marked by a star which is clearly in the normal star forming region of the diagram. }
   \label{fig:bpt}
\end{figure}

As HiZELS is an emission line survey, some fraction of the galaxy sample will be AGN dominated. The ratios of [NII]/$\rm H{\alpha}$ and [OIII]/$\rm H{\beta}$ lines are plotted against each other in a BPT (Baldwin, Phillips \& Terlevich) diagram \citep{baldwin1981} to assess the AGN fraction of the HiZELS sample at $z=1.47$ (Fig. \ref{fig:bpt}). As stated in \S\ref{sec:stat} there are only \totalBPT\ galaxies with all four $\rm H{\alpha}$, $\rm H{\beta}$, [NII] and [OIII] lines present above their detection thresholds. In order to increase the number of galaxies we also include upper limit data points from those with three lines either $\rm H{\alpha}$, $\rm H{\beta}$ and [OIII] (13 galaxies) or $\rm H{\alpha}$, $\rm H{\beta}$ and [NII] (7 galaxies) with the upper limits defined by the detection threshold for the emission lines (S/N=2). Finally, we include 15 galaxies with no detected $\rm H{\beta}$ emission for which we have estimated $\rm H{\beta}$ through the $\rm H{\alpha}$ flux assuming $A_{\rm H\alpha}=1$ and Case B as above. 

We assess the AGN content by considering the positions of the galaxies in the BPT diagram relative to the \cite{kew2001} line, often used as a demarcation between AGN (above) and starbursts (below). For our BPT sample of HiZELS-FMOS galaxies at $z=1.47$, which include those with all four lines, upper limits and estimated $\rm H{\beta}$, $8\%\pm5$ are potential AGN in the \cite{kew2001} definition. If we look at the entire \totalHaNIItwosig\ galaxies with detections of both $\rm H{\alpha}$ and [NII] a further two galaxies have a flux ratio $\rm \log [NII]/\rm H{\alpha}>0.2$, which means they are most probably AGN in the \cite{kew2001} definition (see \S\ref{sec:resid}), giving a total AGN fraction of $8\%\pm3$. We note that the median BPT position of this sample, marked on Fig. \ref{fig:bpt}, is well within the region of the diagram occupied by star-forming galaxies. In other independent analyses of the HiZELS survey, \cite{garn2010a} and \cite{sobral2013} estimate that $5-15\%$ of the $z=0.84$ and $\sim15\%$ of the $z=1.47$ galaxies are AGN through their observed spectral energy distributions and emission line ratios, consistent with our estimate. As we do not have the required spectral line information to assess the AGN content of each individual galaxy and the potential AGN identified are only a small fraction of the total, which lie very close to the \cite{kew2001} line, we do not exclude them from our analysis. We discuss the effect of low level AGN activity on our metallicity analysis in \S\ref{sec:sfrmassmet}.

\subsection{Stellar mass, SFR and metallicity}
\label{sec:sfrmassmet}
To assess the presence of the Fundamental Metallicity Relation at $z\sim0.84-1.47$ we need to obtain reliable estimates of the mass, star formation rate and metallicity for the galaxies in the HiZELS-FMOS sample. The stellar masses are computed by fitting SEDs to the rest-frame UV, optical and near-infrared data available ($FUV$, $NUV$, $U$, $B$, $g$, $V$, $R$, $i$, $I$, $z$, $Y$, $J$, $H$, $K$, $3.6\mu \rm m$, $4.5\mu \rm m$, $5.8\mu \rm m$, $8.0\mu \rm m$ collated in \citealt{sobral2013}, see references therein), following \cite{sobral2011} and the reader is referred to that paper for more details. The SED templates are generated with the \cite{bc2003} package using Charlot \& Bruzual (2007, unpublished) models, a \cite{chabrier2003} IMF, and an exponentially declining star formation history with the form $e^{-t/\tau}$, with $\tau$ in the range 0.1 Gyrs to 10 Gyrs. The SEDs were generated for a logarithmic grid of 200 ages (from 0.1 Myr to the maximum age at each redshift being studied). Dust extinction was applied to the templates using the \cite{calzetti2000} law with $E(B-V)$ in the range 0 to 0.5 (in steps of 0.05), roughly corresponding to A$_{\rm H\alpha}\sim0-2$. The models are generated with different metallicities, including solar; the reader is referred to \cite{sobral2011} and Sobral et al. (in prep.) for further details. For each source, the stellar mass is computed as the median of stellar masses of the solutions which lie within $1\sigma$ of the best fit.

The star formation rates for the HiZELS-FMOS sample are calculated from the aperture-corrected FMOS $\rm H{\alpha}$ luminosity and the relation of \cite{kennicutt1998} corrected to a \cite{chabrier2003} IMF [${\rm SFR ( M_{\odot}yr^{-1})}=4.4\times10^{-42}L_{H\alpha} \rm(erg \,s^{-1})$], assuming a dust extinction $A_{\rm H{\alpha}}=1$\,mag \citep[see \S\ref{sec:dustagn} for Balmer decrement analysis and][]{sobral2013}.

The gas phase abundance of Oxygen [$\rm 12+\log(O/H)$] for the sample can be estimated from the ratio of the [NII] to $\rm H\alpha$ lines \citep{Alloin79, denicolo2002,kewley2002}. This is often referred to as the N2 method, where

\begin{equation}
{\rm N2}=\log (f_{\rm [NII]}/f_{\rm H_\alpha})
\label{eq:n2}
\end{equation}

The median value of N2 for our sample (including the upper limits, assuming [NII] value is that of the S/N=2 detection threshold) is $0.32\pm0.03$. To convert from N2 to Oxygen abundance we use the conversion of \cite{Pettini2004}, which is appropriate for high redshift star-forming galaxies, where:

\begin{equation}
12+\log(\rm O/H)=8.9+0.57\times\rm N2
\label{eq:pet}
\end{equation}

The median metallicity of the HiZELS-FMOS sample, for those with detected [NII], is found to be $\rm 12+\log(O/H)=8.71\pm0.03$. If we include the 44 non-detections of [NII] not affected by the OH sky lines (the 30\% of non-detections discussed in \S\ref{sec:stat}), then this median metallicity falls by 0.09\,dex to $\rm 12+\log(O/H)=8.62\pm0.02$. These values are in agreement with the $z=1.47$ and $z=0.84$ $\rm H\alpha$ emitter stacks featured in Fig. \ref{fig:HaNII}, where $12+\log\rm(O/H)=8.63\pm0.02$ and $8.65\pm0.02$ respectively and all are consistent with the Solar value of $8.66\pm0.05$ \citep{asplund2004}. 

The presence of unaccounted-for AGN may act to bias our metallicities to higher values due to their enhanced N2 values, with both \cite{wright2010} and \cite{Newman2013} finding that at $z>1$ the region of the BPT diagram at the boundary between star-forming galaxies and AGN contains some composite systems with spatially concentrated AGN imbedded within a star-forming galaxy. However, we note that in the case of \cite{Newman2013}, their typical $\rm [OIII]/H\beta$ values are significantly higher ($\rm \log [OIII]/H\beta\sim0.5$) compared to our median ($\rm \log [OIII]/H\beta\sim0$, see Fig. \ref{fig:bpt}), which implies a higher AGN contribution to their sample. We can quantify the effect of this potentially hidden AGN contamination by performing simple cuts in the N2 value regardless of the $\rm [OIII]/H\beta$ ratio. If we remove all galaxies with $\rm N2>-0.2$ then the median metallicity of our sample becomes $8.59\pm0.02$, leaving the result relatively unaffected. However, if we perform a more extreme cut at $\rm N2=-0.3$ then this value falls to $8.55\pm0.02$. This simple cut assumes that all galaxies with N2 ratios in excess of these values are AGN but if we instead randomly removed 50\% of the galaxies with $N2>-0.2$ (-0.3) then the median metallicity (averaged over 1000 resamplings) is $8.59\pm0.02$ ($8.58\pm0.02$). These values agree with our median metallicity for the full sample to within $\sim1\sigma$ and therefore we conclude that our results are robust to the presence of hidden AGN.

\section{Results}
\label{sec:results}
\subsection{The Mass-Metallicity Relation}
\label{sec:mmr}

\begin{figure*}
   \centering
\includegraphics[scale=0.75]{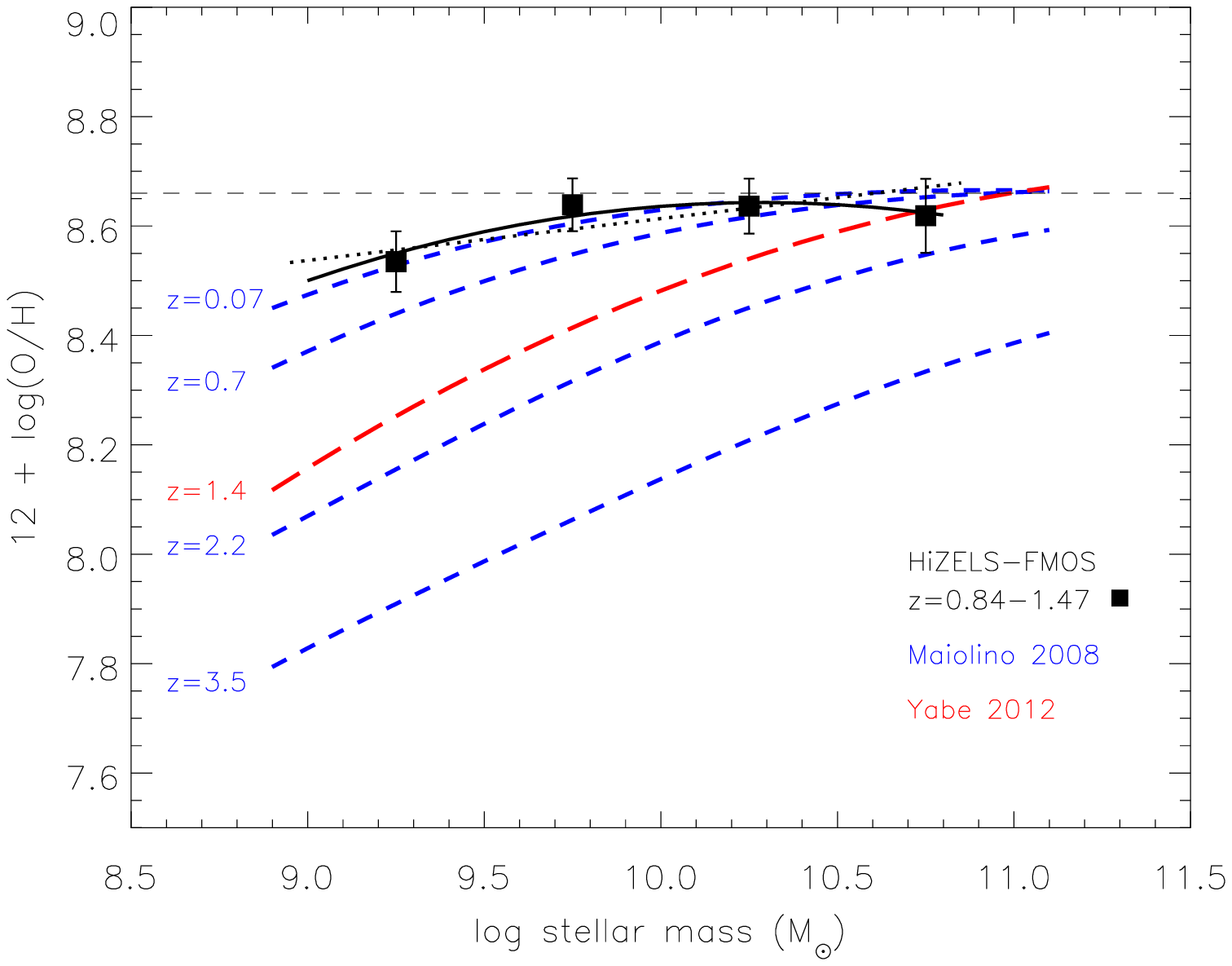} 

\caption[]{The median mass-metallicity relation for our combined $z=0.84-1.47$ HiZELS-FMOS sample (black squares), including the upper limits from the non-detections of [NII], in bins of mass with {\it standard error} bars shown. The solid and dotted black lines are 2nd order (see Eq. \ref{eq:m08}) and linear fits to this relation. The dashed blue lines are the \cite{maiolino2008} fits to the: \cite{kewley2008}, $z=0.07$; and the \cite{savaglio2005}, $z=0.7$; \cite{erb2006}, $z=2.2$; and their own $z=3.5$ datasets corrected to a a \cite{chabrier2003} IMF and to the \cite{Pettini2004} metallicity calibration, for consistency with our data. The long-dash red line is taken from the \cite{yabe2012} sample of $z=1.4$ star-forming  galaxies, which is most relevant to our sample, corrected to a \cite{chabrier2003} IMF. The black dashed horizontal line is the solar value of 8.66 \citep{asplund2004}. The HiZELS-FMOS mass-metallicity relation is in remarkable agreement with the `local', $z=0.07$ SDSS relation of \cite{kewley2008} and is systematically more metal-rich than other samples, which we conclude is due to the significantly higher SFR probed in those studies, which we illustrate in Fig. \ref{fig:massmetsfr}.}
   \label{fig:massmet}
\end{figure*}

The mass-metallicity relation for our combined sample of $z=0.84$ and $z=1.47$ HiZELS-FMOS galaxies is plotted in Fig \ref{fig:massmet}, along with similar studies for comparison. We plot the median metallicity values for the sample (including the upper limits from the [NII] non-detections, see \S\ref{sec:stat}) in bins of 0.5\,dex in mass with their associated standard errors. We also produce median stacks by combining the detected and [NII] upper limit spectra for each bin in mass and find that the resultant mass-metallicity relation is entirely consistent with the median values. The HiZELS-FMOS mass ranges  and median metallicity values from Fig. \ref{fig:massmet} and those from the median stacks are presented in Table \ref{tab:massmet}. We include a fit to the medians of the combined HiZELS-FMOS detected [NII] data and the upper limits of the form: 

\begin{equation}
12+\log(\rm O/H)=-0.0864\,(\log M_{\star}-\log M_{0})^{2} + K_{0}
\label{eq:m08}
\end{equation}

\noindent as used by \cite{maiolino2008} to describe the mass-metallicity relations in their study of $z\sim0.1-3.5$ galaxies (although we note that in their paper they use a \cite{salpeter1955} IMF and their own metallicity calibration). The best fit values are $\log M_{0}=10.29\pm0.31$ and $K_{0}=8.64\pm0.03$. We also perform a linear fit to our data of the form:

\begin{equation}
12+\log(\rm O/H)=\alpha(\log M_{\star})+\beta
\label{eq:massmetlin}
\end{equation}

\noindent which yields $\alpha=0.077\pm0.050$ and $\beta=7.85\pm0.05$. We compare the HiZELS-FMOS fits to the: \cite{kewley2008}, $z=0.07$; \cite{savaglio2005}, $z=0.7$; \cite{erb2006}, $z=2.2$; and the \cite{maiolino2008} $z=3.5$ dataset, which appear to be progressively lower in metallicity with increasing redshift. For consistency with our analysis, the masses are corrected to a \cite{chabrier2003} IMF and to the \cite{Pettini2004} metallicity calibration, using the equations from \cite{Pettini2004} and \cite{maiolino2008}. From this we can see that our results are in remarkable agreement with the `local', $z=0.07$ SDSS relation of \cite{kewley2008}, which is very similar to the SDSS study of \cite{tremonti2004}. Our results are therefore systematically higher in metallicity than the $z=0.7-3.5$ studies of \cite{savaglio2005,erb2006} and \cite{maiolino2008} showing no evolution in redshift for the mass-metallicity relation of the star forming population. 

The most relevant study to compare with our data is \cite{yabe2012}, which is a sample of photometrically selected $z=1.4$ galaxies, albeit with significantly higher SFR than the HiZELS-FMOS sample $>20\rm M_{\odot}yr^{-1}$. The \cite{yabe2012} line is again systematically lower than the HiZELS-FMOS line by $\sim0.3$\,dex at $10^{10}\rm M_{\odot}$. The majority of our HiZELS-FMOS sample are at $z=1.47$ so this discrepancy may be explained by the lower star formation rates probed by our sample. This is in line with the FMR work of \cite{mannucci2010} who find that even high redshift galaxies sit on the local relation once their enhanced SFR, due in part to both selection effects and the rise in the typical sSFR with redshift, has been taken into account. The presence of the FMR at $z=0.84 - 1.47$ is investigated in \S\ref{sec:fmr}. We note that the slope of the mass-metallicity relation for our sample is also flattened relative to that of the comparison studies at similar $z$, which is discussed further in \S\ref{sec:fmr}.

\begin{table}
\begin{center}
\caption[]{The mass-metallicity relation data for the HiZELS-FMOS sample which includes the [NII] upper limits. The stellar mass range corresponds to the bin width and the metallicities are given as both the median values with their associated standard errors and the metallicity calculated from median stacking of all galaxies in the mass range. The number of galaxies per bin is also displayed.}

\label{tab:massmet}
\small\begin{tabular}{lccc}
\hline
$\rm \log(M_{\star}/M_{\odot})$ &Median&Stack& \\
&{$\rm 12+\log(O/H)$} &{$\rm 12+\log(O/H)$} & N. gal\\
\hline
$9.0-9.5$			&$8.53\pm0.05$ &$8.49\pm0.02$&19\\
$9.5-10.0$		&$8.64\pm0.05$ &$8.66\pm0.02$&35\\
$10.0-10.5$		&$8.64\pm0.05$ &$8.63\pm0.02$&32\\
$10.5-11.0$		&$8.62\pm0.07$ &$8.58\pm0.02$&17\\
\hline
\end{tabular}
\end{center}
\end{table}

\subsection{The Fundamental Metallicity Relation}
\label{sec:fmr}

\begin{figure*}
   \centering
\includegraphics[scale=0.75]{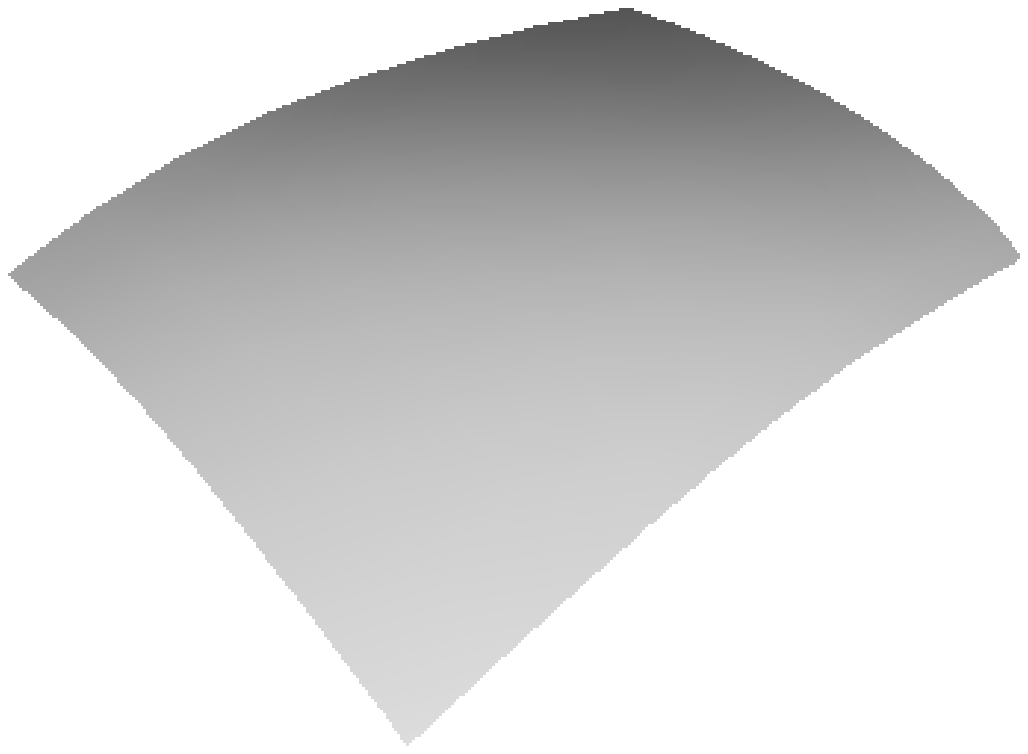} 
\includegraphics[scale=0.5]{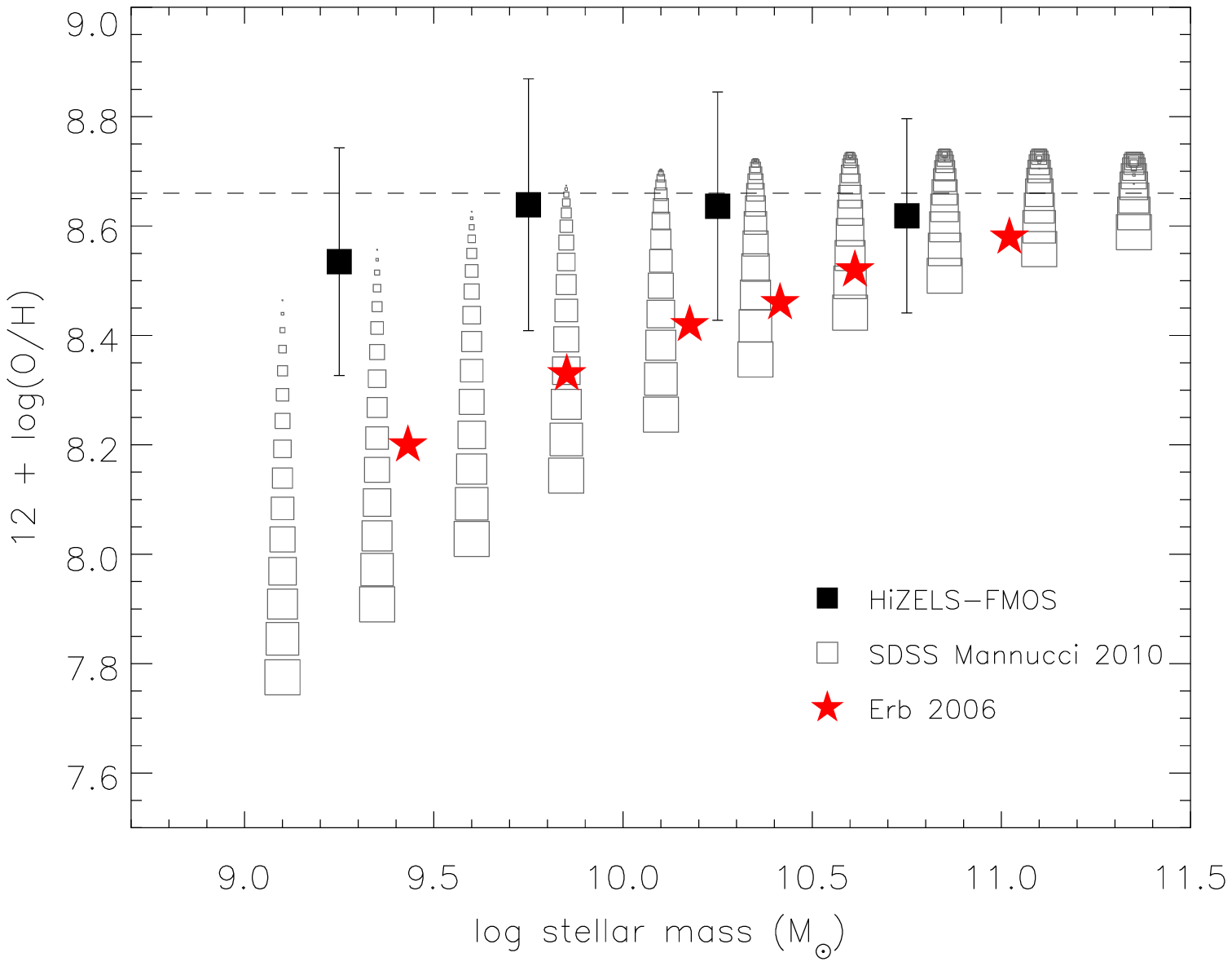} 
\includegraphics[scale=0.5]{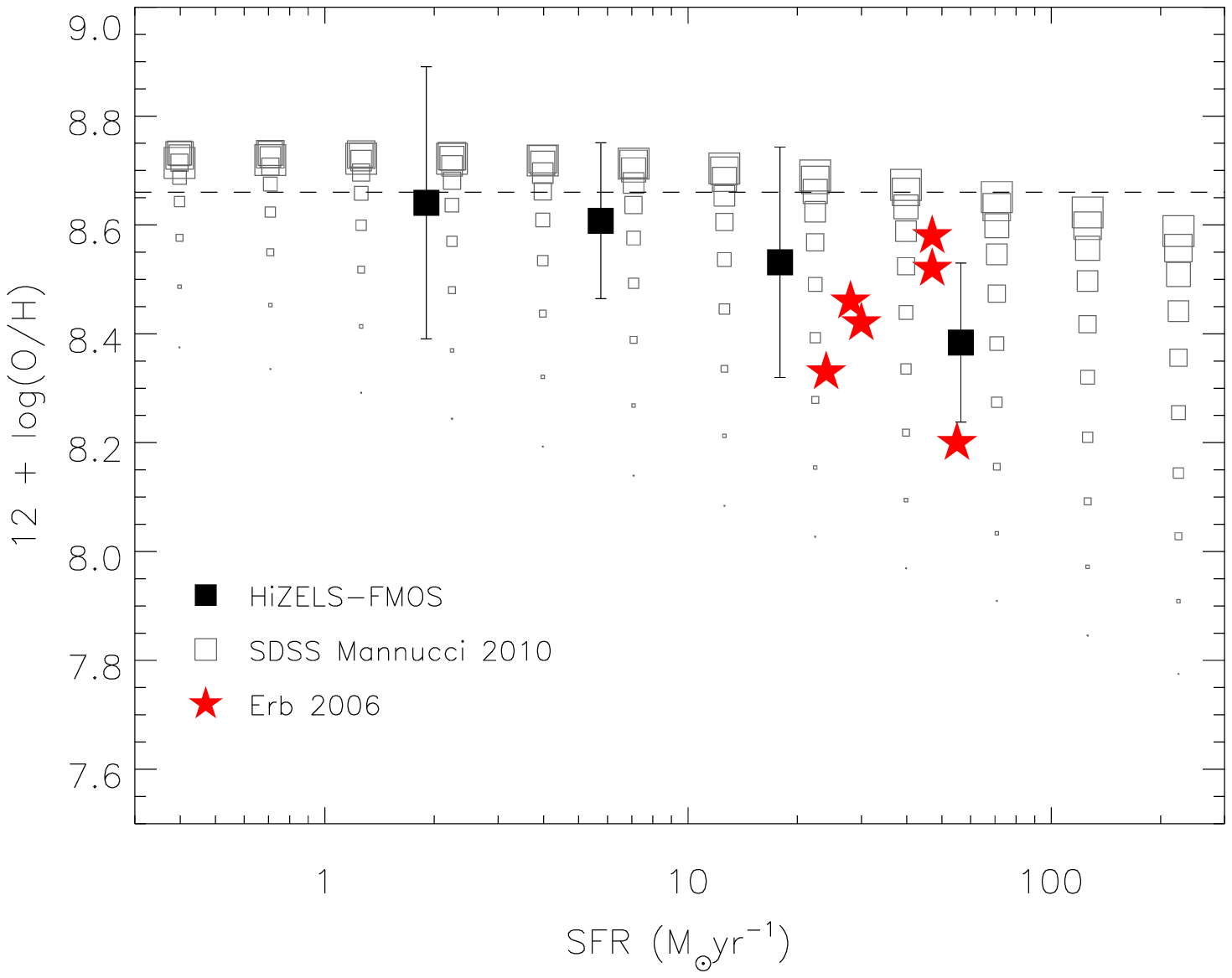} 
\caption[]{{\it Upper:} The three dimensional stellar mass -- metallicity -- SFR plane. The blue filled triangles and black filled circles are the HiZELS-FMOS data at $z=0.84$ and $z=1.47$ respectively. The grey surface represents the low redshift SDSS FMR surface of \cite{mannucci2010}, M10, which we have converted to a \cite{Pettini2004} metallicity calibration. The red stars are the median values from \cite{erb2006} at $z\sim2$. {\it Lower Left:} A projection of the three dimensional plane to show just the mass -- metallicity relation. In this plot the black filled squares and their associated scatter ({\it one standard deviation}) show the median metallicity values per mass bin (including those derived from the [NII] upper limits). The size of the open grey squares representing the \cite{mannucci2010} FMR corresponds to their SFR, with the smallest and largest squares having $\rm \log SFR=-0.9$ and $2.35$ respectively, with each consecutive square size separated by 0.25\,dex. The black dashed horizontal line is the solar value of 8.66 \citep{asplund2004}. {\it Lower Right:} A projection of the three dimensional plane to show just the SFR -- metallicity relation. Again the black filled squares and their associated scatter are the median metallicity values per SFR bin. The size of the open grey squares representing the \cite{mannucci2010} FMR corresponds to their stellar mass, with the smallest and largest squares having $\log \rm M_{\star}=9.0$ and $11.25$ respectively, with each consecutive square size separated by 0.25\,dex. This figure demonstrates that the HiZELS-FMOS galaxies are in broad agreement with \emph{both} the $z\sim0.1$ plane of \cite{mannucci2010} {\it and} the $z=2$ observations of \cite{erb2006} due to the shape of the FMR.}
   \label{fig:massmetsfr}
\end{figure*}

In this section we combine the mass, metallicity and SFR of the galaxies in our HiZELS-FMOS sample to investigate whether a plane such as the FMR of \cite{mannucci2010} exists at $z\sim1-1.5$ in order to discover whether a balance of gas inflow and outflow has already been set up at this redshift. Fig. \ref{fig:massmetsfr} displays the mass-metallicity-SFR information for the HiZELS-FMOS sample. We include the SDSS derived FMR of \cite{mannucci2010}, corrected to the \cite{Pettini2004} metallicity calibration using the equations from \cite{Pettini2004} and \cite{maiolino2008}, for consistency with HiZELS-FMOS (see Table \ref{tab:fmr}). Also plotted are the $z\sim2$ observations of \cite{erb2006} which are an excellent comparison sample to illustrate the argument of \cite{mannucci2010}. The top panel of Fig. \ref{fig:massmetsfr} is a three-dimensional representation of this plane to display its general form, with the FMOS observations as individual points. For ease of comparison we also include the two lower plots which are the mass-metallicity ({\it left}) and the SFR-metallicity ({\it right}) projections of the FMR, with the median values of the HiZELS-FMOS data (including the upper limits due to the non-detections of [NII]) plotted with their associated standard deviations to represent the scatter (c.f. the standard errors in Fig. \ref{fig:massmet}). As with Fig. \ref{fig:massmet} the  {\it lower left} panel shows that the mass metallicity relation of the HiZELS-FMOS sample lies above that of \cite{erb2006} but both lie within the span of values we include from the \cite{mannucci2010} FMR. However, the SFR-metallicity relation ({\it lower right panel}) shows that in fact the \cite{erb2006} sample is highly star-forming and is now in agreement with the HiZELS-FMOS sample. We perform a linear fit to the negative trend between SFR and metallicity the form:

\begin{equation}
12+\log(\rm O/H)=\gamma(\log SFR)+\delta
\label{eq:massmetlin}
\end{equation}

\noindent which yields $\gamma=-0.175\pm0.066$ and $\delta=8.73\pm0.06$. 

From Fig. \ref{fig:massmetsfr} we can see that the HiZELS-FMOS galaxies share the parameter space with both the low redshift FMR of \cite{mannucci2010} and the $z\sim2$ observations of \cite{erb2006}. This FMOS observed, HiZELS sample therefore bridges the gap between the SDSS and $z\sim2$ observations and demonstrates evidence of an FMR in place at $z\sim1.5$.

\subsubsection{Residuals about the FMR}
\label{sec:resid}
\begin{figure*}
   \centering
\includegraphics[scale=0.55, trim=0 60 20 0, clip=true]{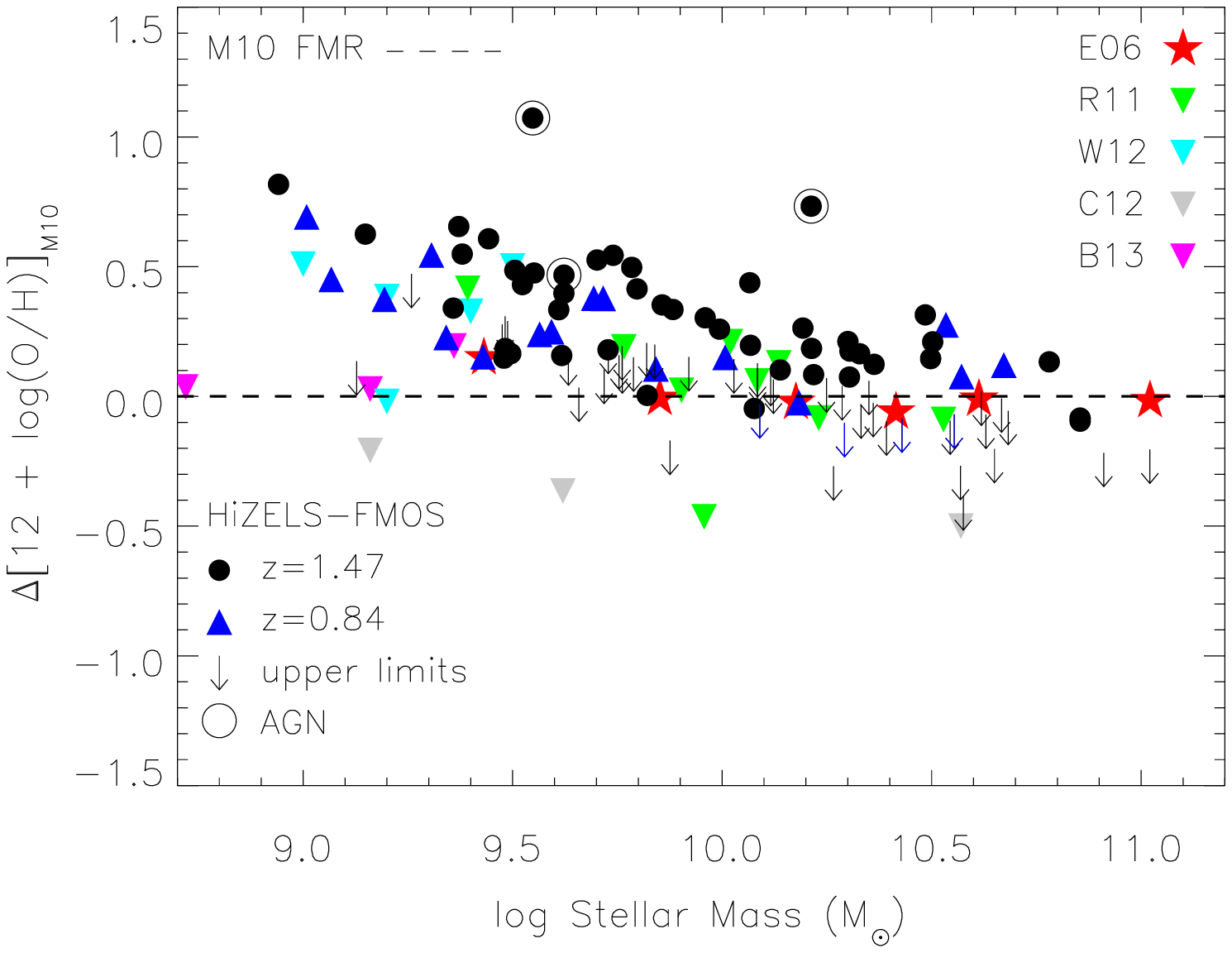} 
\includegraphics[scale=0.55, trim=80 60 0 0, clip=true]{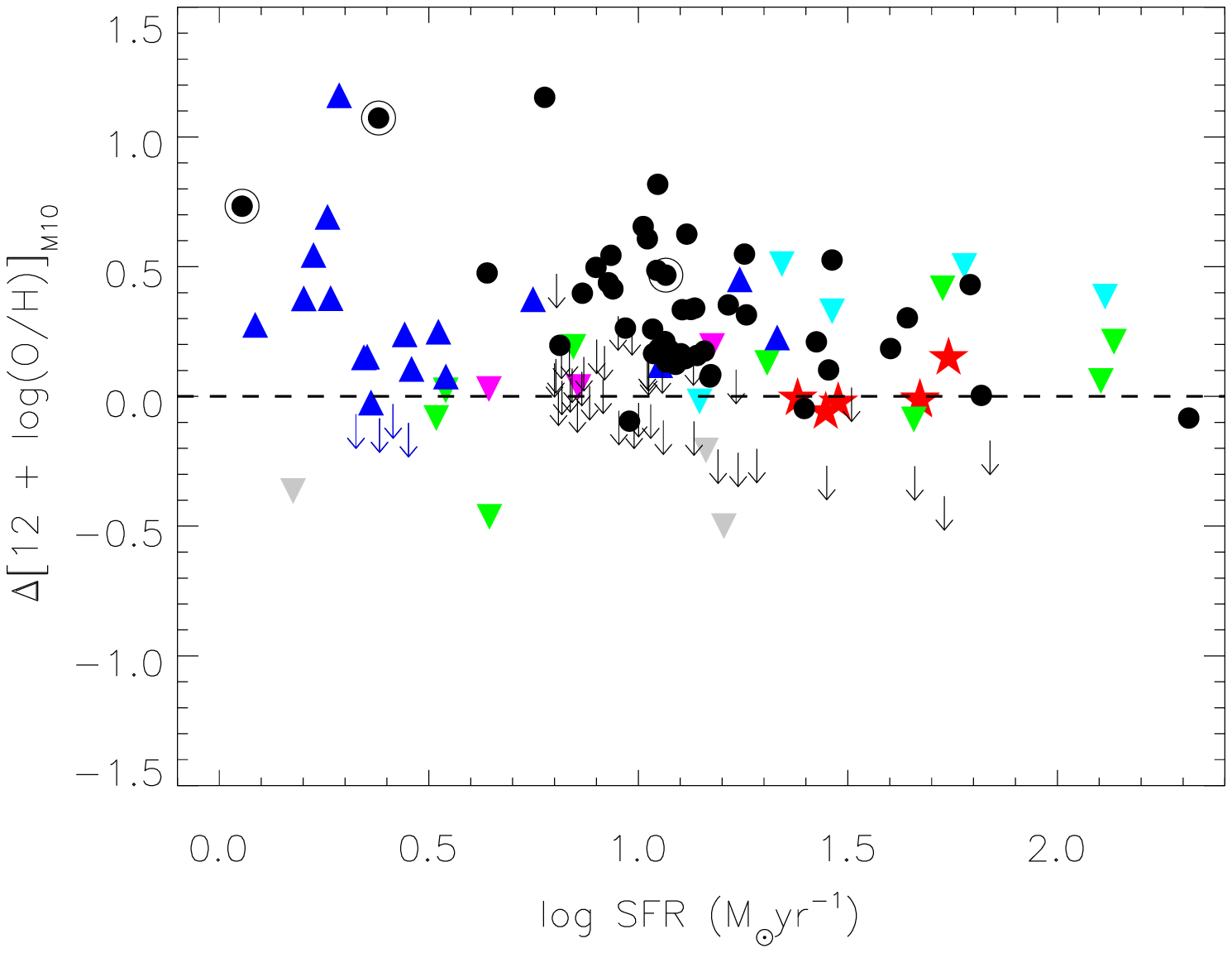} 
\includegraphics[scale=0.55, trim=0 0 20 30, clip=true]{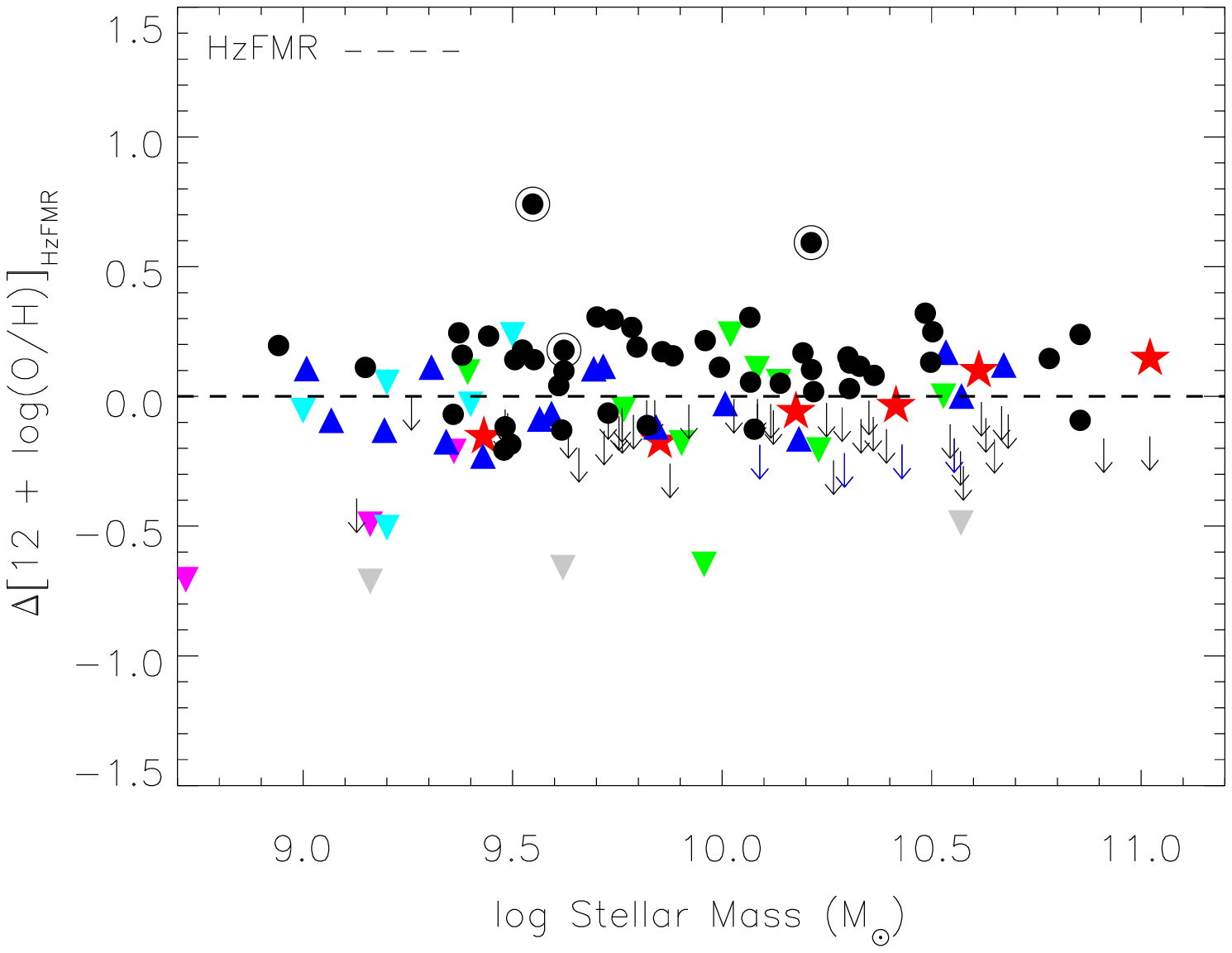} 
\includegraphics[scale=0.55, trim=80 0 0 30, clip=true]{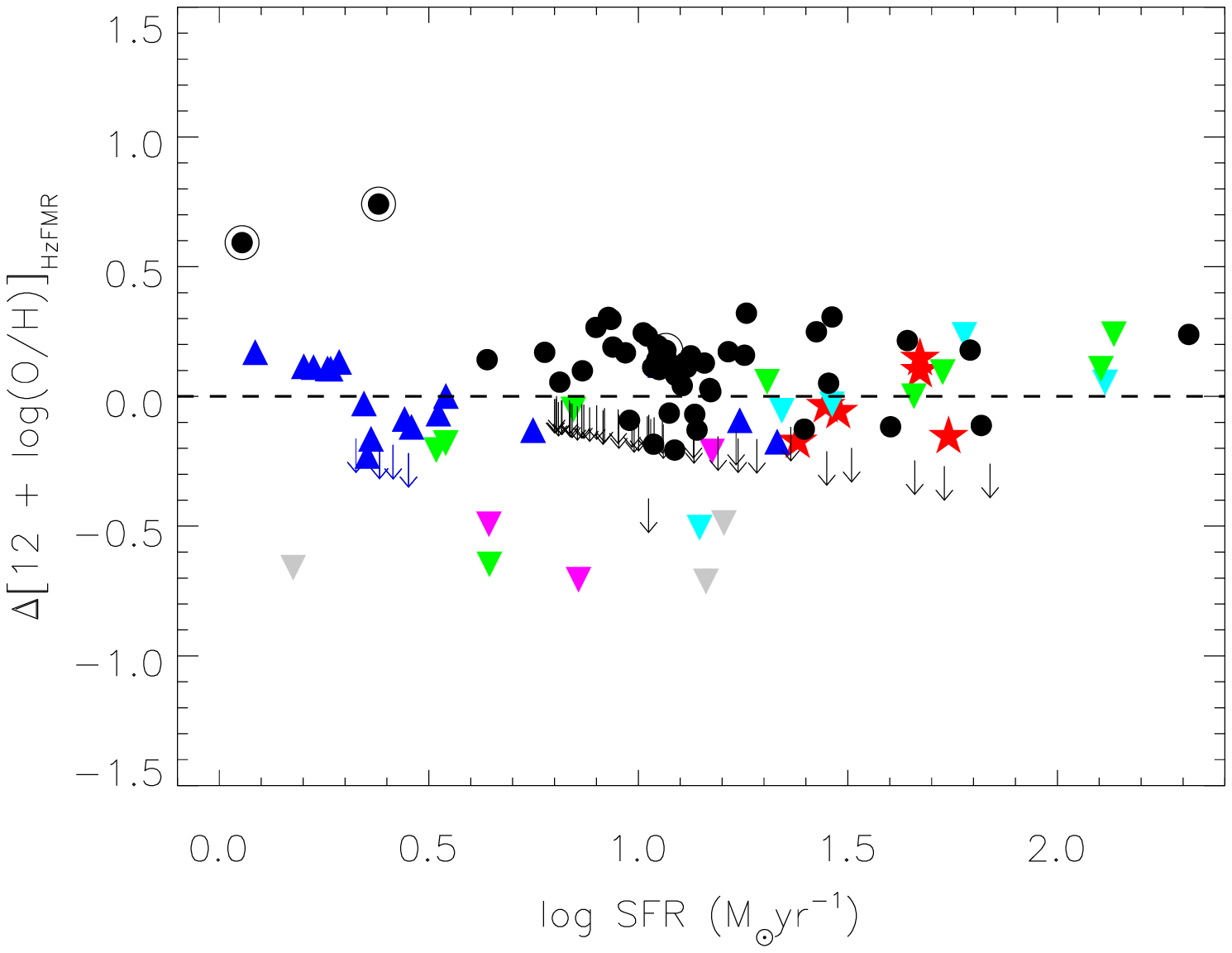} 
\caption[]{\emph{Upper}: The difference between measured metallicity and that predicted by the FMR of \cite{mannucci2010} ($\Delta\rm[12+\log(O/H)]_{M10}$) plotted against stellar mass (left) and SFR (right). Positive values are metallicities higher than that expected by the FMR. The HiZELS-FMOS points are represented by black circles ($z=1.47$) and blue triangles ($z=0.84$). The HiZELS-FMOS galaxies are found to scatter around the $z\sim0.1$ FMR although there is a negative trend between $\Delta\rm[12+\log(O/H)]_{M10}$ and stellar mass.  Open circles denote likely AGN based on the BPT diagram or those with $\log(\rm [NII]/H\alpha)>0.2$ (see Fig. \ref{fig:bpt}). Downward arrows are upper limits for galaxies without detected [NII]. We include the median values of \cite{erb2006} (red stars) and the individual lensed galaxies from \cite{richard2011,wuyts2012,christensen2012} and \cite{belli2013} represented by green, cyan, grey and magenta inverted triangles respectively. \emph{Lower}: The difference between measured metallicity and that predicted by our best-fit HzFMR plane to the HiZELS-FMOS data ($\Delta\rm[12+\log(O/H)]_{HzFMR}$, see Eq. \ref{eq:hizsurf}) plotted against stellar mass ({\it left}) and SFR ({\it right}). The negative trend between $\Delta\rm[12+\log(O/H)]$ is now removed and the scatter significantly reduced.}
   \label{fig:resid}
\end{figure*}

To further demonstrate how the HiZELS-FMOS data compare to the SDSS FMR, in the upper two panels of Fig. \ref{fig:resid} we plot the difference between the measured metallicity of the HiZELS-FMOS galaxies and the metallicity predicted by the $z\sim0.1$ FMR (using the stellar masses and SFRs of our HiZELS-FMOS sample in the \cite{Pettini2004} version of the \citealt{mannucci2010} FMR, see Table \ref{tab:fmr}), hereafter $\Delta\rm [12+\log(O/H)]_{M10}$. In the {\it upper left} panel of Fig. \ref{fig:resid} $\Delta \rm[12+\log(O/H)]_{M10}$ is plotted against stellar mass and in the {\it upper right} against SFR. From this plot one can see that the HiZELS-FMOS observations lie close to the $z\sim0.1$ FMR with the median $\Delta\rm[12+\log(O/H)]_{M10}$ being $0.04$ with a scatter of 0.4\,dex, showing good evidence that there is such a plane in place at $z=1-1.5$. We note that the true scatter may be larger as we have included the upper limits. The scatter around the FMR is significantly larger than that seen at $z\sim0.1$ by \cite{mannucci2010} of 0.05\,dex but this is in part due to the observed negative trend between $\Delta\rm[12+\log(O/H)]_{M10}$ and the stellar mass. Also included on this plot are the binned values of the \cite{erb2006} sample and the individual values from the lensed galaxy samples of \cite{richard2011,wuyts2012,christensen2012} and \cite{belli2013} at $z\sim1.0-3.0$ with $\rm\log (M_{\star}/M_{\odot})>8.5$ and $\rm\log (SFR [M_{\odot}\,yr^{-1}])>0.1$ to match the HiZELS-FMOS sample. These are all converted to the \cite{Pettini2004} metallicity calibration using the equations from \cite{Pettini2004} and \cite{maiolino2008} and, where required for \cite{richard2011}, a \cite{chabrier2003} IMF for the mass and SFR.

Although a negative trend exists between $\Delta\rm[12+\log(O/H)]_{M10}$ and stellar mass, we note that there is no trend with SFR in Fig. \ref{fig:resid}. From the {\it upper left} panel of Fig. \ref{fig:resid}, taken in combination, we see a similar trend between $\Delta\rm[12+\log(O/H)]_{M10}$ and mass in the FMR residuals of the \cite{erb2006,richard2011,wuyts2012,christensen2012} and \cite{belli2013} samples. In their analysis, \cite{belli2013} note the discrepancy between their data and the original $z\sim0.1$ FMR of \cite{mannucci2010} and act to remove it by comparing to the low mass regime version of the FMR presented in \cite{mannucci2011} instead. However, when we compare the HiZELS-FMOS sample to the \cite{mannucci2011} FMR, the trend between $\Delta\rm[12+\log(O/H)]$ and stellar mass is reduced but still remains with a scatter of 0.3\,dex. It therefore appears that the trend between $\Delta\rm[12+\log(O/H)]_{M10}$ and the mass component of the FMR is driven by the observed flattening in the mass-metallicity relation, at all star formation rates, for the HiZELS-FMOS sample (see Fig. \ref{fig:massmetsfr}).

To describe the plane in mass, metallicity and SFR we observe for the HiZELS galaxies at $z=0.84-1.47$ and remove the trend between $\Delta\rm[12+\log(O/H)]$ and stellar mass, we perform a fit to the FMOS data of the same form as \cite{mannucci2010}, that is:

\begin{equation}
12+\log(O/H)=a_0+a_1\,m+a_2\,s + a_3\,m^2 + a_4\,ms + a_5\,s^2
\label{eq:hizsurf}
\end{equation}

\noindent where $m=\log \rm M_{\star}-10$ and $s=\log \rm SFR$. This is a fit to only 108 $z=0.84-1.47$ galaxies (as we include both the 64 [NII] detected galaxies and the 44 upper limits), unlike the $\sim140,000$ galaxies used for calculating the \cite{mannucci2010} FMR, hence we force the parameters $a_i$ to have the same sign as those of \cite{mannucci2010} to maintain a similar surface form. The best fit parameters are given in Table \ref{tab:fmr}. Forcing the parameters to keep the same sign as their \cite{mannucci2010} equivalents means that $a_1$ ($a_3$) becomes zero in the fitting process which, as it was positive (negative) in the \cite{mannucci2010} FMR, is its minimum (maximum) allowed value. Hereafter, we refer to the fit to our HiZELS-FMOS data as the high redshift FMR (HzFMR).

The lower two panels of Fig. \ref{fig:resid} show the residuals in metallicity about this new HzFMR plane, $\Delta\rm[12+\log(O/H)]_{HzFMR}$. For the HiZELS-FMOS data, the trend between the residuals and the stellar mass is now removed and the scatter about this plane is reduced significantly from 0.4\,dex to 0.2\,dex, when compared to the HiZELS-FMOS residuals around the \cite{mannucci2010} FMR. The scatter is broadly consistent with the measurement errors which are typically 0.06\,dex but are in the range $0.01- 0.2$\,dex so there is no strong evidence that this scatter is driven by trends with additional galaxy properties. We note that the trend between $\Delta\rm[12+\log(O/H)]$ and stellar mass is also removed from the comparison samples of \cite{richard2011} and \cite{wuyts2012} with their combined scatter falling from 0.3\,dex for the \cite{mannucci2010} FMR to 0.2\,dex for our HzFMR plane, despite being excluded from the fitting process.

\begin{table}
\begin{center}
\caption[]{The parameters used in Equation \ref{eq:hizsurf} to fit a plane in mass, metallicity and SFR. The HzFMR is a plane fit to the HiZELS-FMOS data, including the upper limits from the non-detections of [NII]. For comparison we also include the parameters we find from a fit to the \cite{mannucci2010} $z\sim0.1$ SDSS FMR after we convert it to a \cite{Pettini2004} metallicity calibration. }

\label{tab:fmr}
\small\begin{tabular}{lcccccc}
\hline
Plane & \multicolumn{6}{c}{fit parameters}  \\
&  $a_0$&$a_1$&$a_2$&$a_3$&$a_4$&$a_5$ \\
\hline
HzFMR&8.77&0.00&-0.055&0.000&0.019&-0.101\\
SDSS FMR$_{\rm PP04}$&8.58&0.24&-0.092&-0.127&0.085&-0.034\\
\hline
\end{tabular}
\end{center}
\end{table}

\section{Discussion \& Summary}
\label{sec:disc}
The main result of this paper is that star-forming galaxies at $z=0.84-1.47$ are on average no less metal abundant than galaxies of similar mass and SFR at $z\sim0.1$, contrary to results from earlier studies of the evolution of the mass-metallicity relation (e.g. \citealt{erb2006, maiolino2008,yabe2012,zahid2013}), which were likely driven in part by the higher average star formation rates of their galaxies. In fact the mass-metallicity relation for the HiZELS-FMOS sample is in remarkable agreement with the low redshift relations of \cite{tremonti2004} and \cite{kewley2008}. However, we note that \cite{zahid2012} do see an evolution to lower metallicities at $z=0.8$ compared to local studies, for a rest-frame UV-optical, photometrically selected sample that includes galaxies in a similar SFR range as HiZELS-FMOS. A UV selection tends to bias against metal-rich/dusty galaxies while being complete for very metal-poor galaxies, which may be the reason for this discrepancy. We also find good evidence that a relationship between mass, metallicity and SFR similar to, but not the same as, the FMR described at low redshift by \cite{mannucci2010} is also present at $z=0.84-1.47$. This is the first demonstration of the presence of such a plane {\it at} $z\gtrsim1$ for a well defined sample of star-forming galaxies with a range of SFRs and masses. 

The simple theoretical explanation for the shape of the FMR is that it is the competing effects of chemical enrichment of the gas by the evolving stellar population, star formation driven winds and the inflow of IGM gas (e.g. \citealt{dave2011}). The correlation between mass and gas phase metallicity is explained, to first order, by strong star formation driven winds driving enriched gas out of lower mass galaxies \citep{larson1974}. The anti-correlation between SFR and metallicity is due to the fuel for new star formation being cool, relatively metal-poor gas inflowing from the IGM (e.g. \citealt{finlator2008}). Although our observed metallicities scatter around the $z\sim0.1$ FMR of \cite{mannucci2010} we observe a flattening of the mass-metallicity relation, at all SFR, which leads to greater disagreement with the FMR at high and low masses (Fig. \ref{fig:resid}). We fit our own plane to the HiZELS-FMOS data, the HzFMR, which accounts for this flattening, providing better agreement with our data and similar studies. We speculate that the reason for the observed flattening in the mass-metallicity relation is that at earlier times the FMR is in its infancy so the balance of gas inflow and outflow may have not yet set up the steeper relation at higher SFR. 

We observe a negative correlation between SFR and metallicity in the HzFMR, which is consistent with the picture that the elevated SFRs of typical galaxies at $z\gtrsim1$ are caused by the increased efficiency of metal-poor IGM gas inflow (e.g. \citealt{keres2005,bower2006,dekel2009}). This enhanced gas inflow is the likely cause of the increase in the observed star formation rate density (SFRD) of the Universe \citep{lilly1996,madau1996,sobral2013}, and it is therefore not driven by major mergers (as we also demonstrated through direct observation in \citealt{sobral2009} and \citealt{stott2013}), which may instead act to maintain or enhance the metallicity of the gas, not dilute it.

Through the inclusion of $\rm H\beta$ and [OIII] observations we estimate the dust content and AGN fraction of our HIZELS-FMOS sample. The average dust attenuation in the $V$ band and at the wavelength of $\rm H\alpha$ (6563\AA) are found, through the Balmer decrement, to be $A_V=1.3\pm0.2$ and $A_{\rm H\alpha}=1.1\pm0.2$ respectively. This is in agreement with the $A_{\rm H\alpha}=1$ used throughout this paper and for HiZELS papers such as \cite{sobral2013} and \cite{stott2013}. 

We use the BPT diagram \citep{baldwin1981} to estimate the potential AGN fraction and find it to be consistent with low redshift samples with $\sim10\%$ of the galaxies occupying the region of the BPT diagram found to contain AGN, using the \cite{kew2001} definition, in agreement with independent, multi-wavelength analyses of the HiZELS survey (\citealt{garn2010a} and \citealt{sobral2013}).

In conclusion, this study is the first of its kind to demonstrate the presence of the FMR, or a variant of it (the HzFMR), at $z\gtrsim1$ with a well understood sample of typical star-forming galaxies. This means that the bulk of the metal enrichment for the $z\sim1-1.5$ star-forming galaxy population takes place in the 4\,Gyr before $z=1.5$. The challenge for galaxy evolution theory is to explain how this balance between the competing processes of gas inflow and outflow creates a mass, metallicity and SFR plane by at least $z=1.5$. Taken in concert with our results from \cite{sobral2013} and \cite{stott2013}, that there is little evolution in the mass, size and merger rates of the star-forming galaxy population since $z\sim2$, this lack of change in the gas phase metallicity is further evidence that, \emph{taken as a population}, many of the properties of star-forming galaxies have remained remarkably constant over $\sim9$\,Gyr of cosmic time, despite the significant change in typical SFR. We speculate that the negative slope of the SFR-metallicity component of the HzFMR plane, combined with the lack of evolution in the number density of major mergers for star-forming galaxies \citep{stott2013}, demonstrates that it is the efficient inflow of metal-poor gas from the IGM that leads to this increase in typical SFR with redshift and is ultimately responsible for the rise in the SFRD of the Universe at $z\gtrsim1$.

\vspace{1in}
\noindent{\bf ACKNOWLEDGEMENTS}

The authors wish to thank Kentaro Aoki for technical assistance and Fabrice Durier for useful discussions. JPS, RGB and IRS acknowledge support from the UK Science and Technology Facilities Council (STFC) under ST/I001573/1. DS acknowledges the award of a Veni Fellowship. IRS acknowledges support from a Leverhulme Fellowship, the ERC Advanced Investigator programme DUSTYGAL and a Royal Society/Wolfson Merit Award. TK acknowledges financial support by a Grant-in-Aid for the Scientific Research (Nos.\, 23740144; 24244015) by the Japanese Ministry of Education, Culture, Sports and Science.

Based on data collected at the Subaru Telescope, which is operated by the National Astronomical Observatory of Japan.

The United Kingdom Infrared Telescope is operated by the Joint Astronomy Centre on behalf of the Science and Technology Facilities Council of the UK

\bibliographystyle{mn2e}
\bibliography{fmos_postaccept}

\end{document}